\documentclass{jfm}

\usepackage{color}
\usepackage{graphicx}
\usepackage{epstopdf}
\usepackage{natbib}
\usepackage{bm}

\ifCUPmtlplainloaded \else
  \checkfont{eurm10}
  \iffontfound
    \IfFileExists{upmath.sty}
      {\typeout{^^JFound AMS Euler Roman fonts on the system,
                   using the 'upmath' package.^^J}%
       \usepackage{upmath}}
      {\typeout{^^JFound AMS Euler Roman fonts on the system, but you
                   dont seem to have the}%
       \typeout{'upmath' package installed. JFM.cls can take advantage
                 of these fonts,^^Jif you use 'upmath' package.^^J}%
      }
  \else
  \fi
\fi

\ifCUPmtlplainloaded \else
  \checkfont{msam10}
  \iffontfound
    \IfFileExists{amssymb.sty}
      {\typeout{^^JFound AMS Symbol fonts on the system, using the
                'amssymb' package.^^J}%
       \usepackage{amssymb}%
         \let\leq=\leqslant

         \let\geq=\geqslant
      }{}
  \fi
  \fi
  
\ifCUPmtlplainloaded \else
  \IfFileExists{amsbsy.sty}
    {\typeout{^^JFound the 'amsbsy' package on the system, using it.^^J}%
     \usepackage{amsbsy}}
    {\providecommand\boldsymbol[1]{\mbox{\boldmath $##1$}}}
\fi

\def\Xint#1{\mathchoice
{\XXint\displaystyle\textstyle{#1}}%
{\XXint\textstyle\scriptstyle{#1}}%
{\XXint\scriptstyle\scriptscriptstyle{#1}}%
{\XXint\scriptscriptstyle\scriptscriptstyle{#1}}%
\!\int}
\def\XXint#1#2#3{{\setbox0=\hbox{$#1{#2#3}{\int}$}
\vcenter{\hbox{$#2#3$}}\kern-.5\wd0}}

\def\dashint{\Xint {\boldsymbol -}}

\providecommand\bcdot{\boldsymbol{\cdot}}

\newsavebox{\astrutbox}
\sbox{\astrutbox}{\rule[-5pt]{0pt}{20pt}}

\let\vec\boldsymbol

\newcommand{\eqref}[1]{(\ref{#1})}

\title[Magnetic droplet]{Magnetic micro-droplet in rotating field: numerical simulation and comparison with experiment}

\author[J.Erdmanis, G. Kitenbergs, R. Perzynski, A. C\={e}bers]%
{J.Erdmanis$^{1}$, G. Kitenbergs$^{1}$, \\
R. Perzynski$^{2}$, A. C\={e}bers$^{1,3}$\thanks{Email address for correspondence: aceb@tok.sal.lv}}

\affiliation{
$^1$MMML lab, Faculty of Physics and Mathematics, University of Latvia, Riga, LV-1002, Latvia, \\%
$^2$Sorbonne Universit\'es, UPMC Univ Paris 06, CNRS, UMR 8234, PHENIX, Paris, F-75005, France,\\%
$^3$Chair of Theoretical Physics, University of Latvia, Riga, LV-1002, Latvia

{\bf This is a first revision of the article, submitted to the Journal of Fluid Mechanics (cambridge.org/core/journals/journal-of-fluid-mechanics)}
}

\pubyear{}
\volume{}
\pagerange{}
\date{?; revised ?; accepted ?. - To be entered by editorial office}
\begin{document}

\maketitle

\begin{abstract}
Magnetic droplets obtained by induced phase separation in a magnetic colloid show a large variety of shapes when exposed to an external field. However, the description of shapes is often limited. Here we formulate an algorithm based on three  dimensional boundary-integral equations  for strongly magnetic droplets in a high-frequency rotating magnetic field, allowing us to find their figures of equilibrium in three dimensions. The algorithm is justified by a series of comparisons with known analytical results. We compare the calculated equilibrium shapes with experimental observations and find a good agreement. The main features of these observations are the oblate-prolate transition, the flattening of prolate shapes with the increase of magnetic field strength and the formation of star-fish like equilibrium shapes. We show both numerically and in experiments that the magnetic droplet behaviour may be described with a tri-axial ellipsoid approximation. Directions for further research are mentioned, including the dipolar interaction contribution to the surface tension of the magnetic  droplets, account for the large viscosity contrast between the magnetic droplet and the surrounding fluid.
\end{abstract}
\section{\label{sec:intro}Introduction}
The behaviour of droplets under the action of magnetic fields of different configurations is an important issue in many fields, including microfluidics \citep{1}, mechanics of tissues \citep{2,3}, studies of dynamic self-assembly \citep{23} and many others. After the successful synthesis  of magnetic fluids in the late sixties by \citep{4} an exciting story about droplets of magnetic fluid began. At first the elongation of the droplets  in an external field was observed and explained by \citet{5}. The characteristic hysteresis at a sufficiently  high magnetic susceptibility of the magnetic fluid was described by \citet{6}. This problem is quite similar to the deformation of a conducting liquid droplet studied by \citet{7}, which also described the so-called Taylor cone and the jet formation from the tip of droplet. Recently, progress has been made in studying the dynamics of magnetic drop deformation \citep{24}.

\begin{figure}
\center
\includegraphics[width=1.0\textwidth]{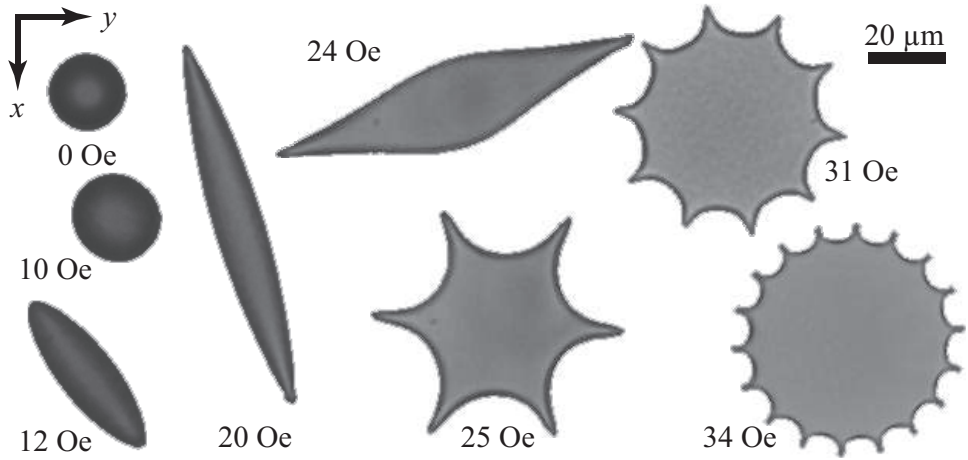}
\caption{Stationary shapes of a single magnetic droplet ($R_0=10~\mu \rm{m}$) in a magnetic field  rotating in the $(x,y)$ plane at various magnetic field $H_{0}$ values. Field frequency is $500~Hz$.}
\label{fig:Rot_ims}
\end{figure}

Phase separation of magnetic colloids enabled researchers to obtain magnetic micro-droplets with a very high magnetic  susceptibility allowing the observation of different new phenomena. One particularly interesting case is the behaviour of a magnetic micro-droplet under the action of a rotating magnetic field \citep{8}. At low  and intermediate rotating field frequencies synchronous and asynchronous rotation of the droplet with respect to the rotating field was observed by \citet{9}. Depending on the magnetic field strength and frequency, various shapes were seen \citep{25}. A family of stationary droplet shapes is observed in the present work as shown in figure~\ref{fig:Rot_ims}. They are formed under the action of a high frequency rotating field. A detailed description of the observed sequence of bifurcations is given below.

In a high frequency field the characteristic time of the shape deformation is much larger than the field period, therefore it is possible to average the magnetic field energy with respect to the field period. As a result the droplet shape is a figure of equilibrium and is determined by the balance of the ponderomotive forces of the self-magnetic field and the capillary forces \citep{8}. On this basis in \citep{8,10} the re-entrant transition for the sequence of shapes  oblate-prolate-oblate was described by applying the virial method and assuming that prolate ellipsoid of revolution arises at the oblate-prolate-oblate bifurcations. Later these results were extended in \citep{11,12} considering the general ellipsoid with three non-equal axes at the oblate-prolate-oblate bifurcations. An essential assumption of these models is the ellipsoidal shape of the droplet.

Our aim is to answer the following question.  Is it possible to describe the set of arising bifurcations by a simple model of magnetic fluid \citep{4}?  For that we develop a numerical algorithm for a droplet in a rotating field which is free from the limitations of the ellipsoidal approximation. We compare the numerical results with experimental ones (including those shown in figure~\ref{fig:Rot_ims}) and also with analytical results, where it is possible. We organize the paper as follows. In \S~\ref{sec:mathform} we formulate the model and give the available theoretical results. The numerical algorithm is based on a boundary integral equation technique. It is described with its tests in \S~\ref{sec:num}. The implementation written in the Julia programming language can be found in \citet{22}. In \S~\ref{sec:res} the numerical analysis of the magnetic droplet shape bifurcations is carried out. The presentation of the experimental results and their comparison with numerical and theoretical results are carried out in \S~\ref{sec:exp}, before a short discussion and conclusion.

\section{\label{sec:mathform}Model and analytical results}
\subsection{\label{sec:2.1}Equation of motion and energy functional}
Equations of motion of  the magnetic fluid reads \citep{4,13}
\begin{equation}
-\nabla p+\frac{\partial \vec{\sigma}^{v}_{k}}{\partial x_{k}}+\frac{\partial \vec{T}_{k}}{\partial x_{k}}=0;~\nabla\bcdot\vec{v}=0~,
\label{Eq:1}
\end{equation}
where $\sigma^v_{ik} = \eta \left( \frac{\partial v_i}{\partial x_k} + \frac{\partial v_k}{\partial x_i} \right)$ is the viscous stress tensor, $\eta$ is the viscosity, $p$ is the pressure, $\vec{e}_{i}\bcdot\vec{T}_{k}=\frac{1}{4\pi}\Bigl(H_{i}B_{k}-\frac{1}{2}H^{2}\delta_{ik}\Bigr)$ is the Maxwell stress tensor and $\frac{\partial \vec{T}_{k}}{\partial x_{k}}=(\vec{M}\bcdot \nabla)\vec{H}$. According to the equations of magnetostatics $\nabla\bcdot\vec{B}=0;~\vec{B}=\vec{H}+4\pi\vec{M}$  and 
$\nabla\times\vec{H}=4\pi\vec{J}/c$ ($\vec{J}$  is the current density in external coils creating the magnetic field and equal to zero throughout the volume of the magnetic fluid.) In the framework of the linear response, magnetization $\vec{M}$ expresses as $\vec{M}=\chi\vec{H}$, with magnetic susceptibility $\chi$ such as $\mu=1+4 \pi \chi = const$, $\mu$ being the magnetic permeability. The equilibrium shapes of the magnetic fluid droplet are usually  analysed considering the energy functional \citep{6} (see \citep{14} for the derivation of the magnetic field contribution)
\begin{equation}
E=-\frac{1}{2}\int\vec{M}\bcdot\vec{H}_{0}dV+\gamma S~,
\label{Eq:2}
\end{equation}
where $\vec{H}_{0}$ is the field strength of external homogeneous  magnetic field, $S$ is  the surface area of a droplet and $\gamma$ is the surface tension (supposed here independent on $H$). Although this approach, as it should be expected, is reasonable,  for the completeness of further analysis we  illustrate below that it directly follows from the dynamic  equations (\ref{Eq:1}) and the boundary conditions describing the evolution of the shapes of the droplet.

Considering the Lagrange variation of some variable $f$ ($\vec{\xi}(\vec{x})$ is the displacement of material element with a radius vector $\vec{x}$)
\begin{equation}
\delta_{L}f=f'(\vec{x}')-f(\vec{x});~\vec{x}'=\vec{x}+\vec{\xi}(\vec{x})
\label{Eq:3}
\end{equation}
and its derivative
\begin{equation}
\delta_{L}\frac{\partial f}{\partial x_{i}}=\frac{\partial \delta_{L}f}{\partial x_{i}}-\frac{\partial \xi_{k}}{\partial x_{i}}\frac{\partial f}{\partial x_{k}}~,
\label{Eq:4}
\end{equation}
while the current of external source is fixed according to (\ref{Eq:4}) we have ($\psi$ being the potential of the magnetostatic field defined by $\vec{H}=\vec{\nabla}\psi$)
\begin{equation}
\delta_{L}H_{i}=\frac{\partial\delta_{L}\psi}{\partial x_{i}}-H_{k}\frac{\partial \xi_{k}}{\partial x_{i}}~.
\label{Eq:5}
\end{equation}
For simplicity we further assume that the fluid is incompressible ($\nabla\bcdot\vec{\xi}=0$).
According to Eq.(\ref{Eq:5}) we  have
\begin{equation}
\frac{1}{4\pi}B_{i}\delta_{L}H_{i}=\frac{1}{4\pi}B_{i}\frac{\partial \delta_{L}\psi}{\partial x_{i}}-\frac{1}{4\pi}H_{i}B_{k}\frac{\partial \xi_{i}}{\partial x_{k}}~.
\label{Eq:6}
\end{equation}
Integrating the relation Eq.(\ref{Eq:6}) over the entire volume and accounting for the boundary conditions on the surface of a droplet $\delta_{L}\psi^{i}=\delta_{L}\psi^{e}$ and $B_{n}^{i}=B_{n}^{e}$ we obtain (by superscripts $^{i}$ and $^{e}$ we denote the values on the inside and outside boundaries of the droplet respectively)
\begin{equation}
\frac{1}{4\pi}\int B_{i}\delta_{L}H_{i}dV=\int\xi_{i}\frac{\partial}{\partial x_{k}}\Bigl(\frac{H_{i}B_{k}}{4\pi}\Bigr)dV-\int\Bigl(\frac{H_{i}^{i}B_{k}^{i}}{4\pi}-\frac{H_{i}^{e}B_{k}^{e}}{4\pi}\Bigr)\xi_{i}n_{k}dS~.
\label{Eq:7}
\end{equation}
Taking into account the identity valid for an incompressible fluid
\begin{equation}
-\int\xi_{i}\frac{\partial}{\partial x_{k}}\Bigl(\frac{H^{2}}{8\pi}\delta_{ik}\Bigr)dV+\int\xi_{i}n_{k}\Bigl(\frac{(H^{i})^{2}}{8\pi}-\frac{(H^{e})^{2}}{8\pi}\Bigr)\delta_{ik}dS=0
\label{Eq:8}
\end{equation}
the relation (\ref{Eq:7}) may be rewritten as follows
\begin{equation}
\frac{1}{4\pi}\int B_{i}\delta_{L}H_{i}dV=\int\xi_{i}\frac{\partial T_{ik}}{\partial x_{k}}dV-\int\xi_{i}n_{k}(T_{ik}^{i}-T_{ik}^{e})dS~.
\label{Eq:9}
\end{equation}
Using $\xi_{i}=v_{i}\delta t$, integrating the dot product of Eq.(\ref{Eq:1}) with the fluid velocity field and accounting for the incompressibility we obtain
($\sigma_{ik}=-p\delta_{ik}+\sigma^{v}_{ik}$)
\begin{equation}
\int v_{i}(\sigma^{i}_{ik}+T^{i}_{ik}-\sigma^{e}_{ik}-T^{e}_{ik})n_{k}dS+\frac{1}{4\pi}\int\vec{B}\bcdot\frac{\delta_{L}\vec{H}}{\delta t}dV-\int\frac{\partial v_{i}}{\partial x_{k}}\sigma^{v}_{ik}dV=0~.
\label{Eq:10}
\end{equation}
The balance of forces on the boundary of the droplet gives 
\begin{equation}
(\sigma^{i}_{ik}+T^{i}_{ik}-\sigma^{e}_{ik}-T^{e}_{ik})n_{k}=-\gamma\Bigl(\frac{1}{R_{1}}+\frac{1}{R_{2}}\Bigr)n_{i}~,
\label{Eq:11}
\end{equation}
where $1/R_{1,2}$ are principal curvatures. 
As a result we conclude that the variation of the sum of the surface energy $\gamma S$ and the thermodynamic potential at fixed current in the external coils $\tilde{F}=-\frac{1}{8\pi}\int\vec{B}\bcdot\vec{H}dV$ equals the amount of the dissipated energy per unit time due to the viscosity of the fluid
\begin{equation}
\frac{d}{dt}(\tilde{F}+\gamma S)=-\int \frac{\partial v_{i}}{\partial x_{k}}\sigma^{v}_{ik}dV~.
\label{Eq:12}
\end{equation}
Instead of the thermodynamic potential $\tilde{F}$, the potential obtained by adding a constant value  $\frac{1}{8\pi}\int H_{0}^{2}dV$ is usually considered \citep{14}, thus leading to  
\begin{equation}
\tilde{F}+\frac{1}{8\pi}\int H_{0}^{2}dV=-\frac{1}{2}\int \vec{M}\bcdot\vec{H}_{0}dV~,
\label{Eq:13}
\end{equation}
and
\begin{equation}
\frac{d}{dt}\Bigl(\gamma S-\frac{1}{2}\int \vec{M}\bcdot\vec{H}_{0}dV\Bigr)=-\int \frac{\partial v_{i}}{\partial x_{k}}\sigma^{v}_{ik}dV~.
\label{Eq:13a}
\end{equation}
The energy $\gamma S-\frac{1}{2}\int \vec{M}\bcdot\vec{H}_{0}dV$,  which is usually minimized to describe the equilibrium shapes of the magnetic droplets \citep{6,13,15}, is in fact decreasing with time due to the viscous dissipation and reaching the minimal value at equilibrium \citep{27}.

The arguments given above show that the figures of equilibrium of the magnetic fluid droplets may be obtained by numerical simulation of the evolution of shapes of a viscous droplet suspended in a surrounding viscous fluid, as described by Eq. (\ref{Eq:1}) and the boundary conditions Eq.(\ref{Eq:11}). For simplicity we will here consider the case  of both the magnetic and the surrounding fluids having equal viscosities. 

\subsection{\label{sec:2.2} Dynamics of droplet shape in an applied field}
We now formulate the set of equations which are solved numerically for the simulation of the droplet shape dynamics in a rotating field. The equation of motion (\ref{Eq:1}) in the case of a constant magnetic susceptibility may be rewritten in the form of the Stokes equation
\begin{equation}
-\nabla\tilde{p}+\eta\Delta\vec{v}=0~,
\label{Eq:20}
\end{equation}
where $\tilde{p}=p-\frac{(\mu-1)\vec{H}^{2}}{8\pi}$. The boundary condition (\ref{Eq:11}) accounting for $T^{(i)}_{nn}-T^{(e)}_{nn}=-2\pi M_{n}^{2}$ and $T^{(i)}_{tn}=T^{(e)}_{tn}$ may be rewritten as follows
\begin{equation}
(-\tilde{p}\delta_{ik}+\sigma^{v(i)}_{ik})n_{k}-(-p\delta_{ik}+\sigma^{v(e)}_{ik})n_{k}=\frac{\mu-1}{8\pi}\vec{H}^{(i)2}n_{i}+2\pi M^{2}_{n}n_{i}-\gamma\Bigl(\frac{1}{R_{1}}+\frac{1}{R_{2}}\Bigr)n_{i}~.
\label{Eq:21}
\end{equation}
From the relation (\ref{Eq:21}) we see that the fluid motion is determined by the action of the normal force on the boundary of the droplet
\begin{equation}
\vec{f}=\frac{\mu(\mu-1)}{8\pi}H^{(i)2}_{n}\vec{n}+\frac{\mu-1}{8\pi}H^{(i)2}_{t}\vec{n}-\gamma\Bigl(\frac{1}{R_{1}}+\frac{1}{R_{2}}\Bigr)\vec{n}~.
\label{Eq:22}
\end{equation}
Its velocity for equal drop and surrounding fluid  viscosities is given as an integral over surface forces as a special case of the Stokes BIE \cite{19}
\begin{equation}
v_{i}(\vec{x})=\frac{1}{8\pi\eta}\int\Bigl(\frac{\delta_{ik}}{|\vec{x}-\vec{y}|}+\frac{(x_{i}-y_{i})(x_{k}-y_{k})}{|\vec{x}-\vec{y}|^{3}}\Bigr)f_{k}(\vec{y})dS_{\vec{y}}~.
\label{Eq:23}
\end{equation}
The normal component of the liquid velocity (\ref{Eq:23}) on the interface gives its position in the next time step.
We see that the dynamics is known if the normal $H^{(i)}_{n}$ and tangential $H^{(i)}_{t}$ components of the magnetic field strength on the internal side of the droplet boundary are found. These components are given by the solution of equations of magnetostatics $\nabla\bcdot\vec{H}=0;~\nabla\times\vec{H}=0$ with boundary conditions $H^{(i)}_{t}=H^{(e)}_{t}$, $\mu H^{(i)}_{n}=H^{(e)}_{n}$ and $\vec{H}(\vec{x})\rightarrow \vec{H}_{0}$ at $|\vec{x}|\rightarrow \infty$.
This problem of magnetostatics may be efficiently solved by applying a boundary integral equation technique which we describe in the next section.

\subsection{\label{sec:2.3} Ellipsoidal figures of equilibrium}
The result (\ref{Eq:13a}) allows us to find some of the simplest figures of equilibrium of the magnetic fluid droplet in a high frequency rotating magnetic field. If the period of the rotating field is much smaller than the characteristic time $\gamma/\eta R_{0}$ of the droplet shape relaxation, where $R_{0}$ is the radius of initial spherical droplet, we may average the magnetic energy with respect to the period of the  rotating field and consider the minimum of the energy functional $\gamma S-\frac{1}{2}\int \vec{M}\bcdot\vec{H}_{0}dV$.
As an approximation we assume that the droplet has the shape of a general ellipsoid with the semi-axes $a,b,c$ and its interface is determined by the equation
$x^{2}/a^{2}+y^{2}/b^{2}+z^{2}/c^{2}=1$.

As the droplet always has an axial symmetry at a small field strength we
consider it here as an oblate ellipsoid of revolution with the semi-axes $a=b>c$. The demagnetizing field coefficients are  $n_{1}=n_{2}$ in the plane $(x,y)$ of the rotating field and $n_{3}$ in the perpendicular direction. As a result the energy of the  droplet reads ($K=a/c>1$)
\begin{equation}
\frac{E}{2\pi\gamma R^{2}_{0}}=K^{2/3}\Bigl(1+\frac{1}{2K^{2}\sqrt{1-1/K^{2}}}\log{\frac{1+\sqrt{1-1/K^2}}{1-\sqrt{1-1/K^2}}}\Bigr)-\frac{H^{2}_{0}R_{0}}{12\pi\gamma}\frac{1}{1/(\mu-1)+n_{1}}~,
\label{Eq:14}
\end{equation}
where $n_{1}=(1-n_{3})/2$ and
\begin{equation}
n_{3}=\frac{K^{2}}{(K^{2}-1)^{3/2}}\Bigl((K^2-1)^{1/2}-\arctan{\sqrt{K^2-1}}\Bigr)~.
\label{Eq:15}
\end{equation}
We obtain the axis ratio $K$ for the equilibrium shape by  minimizing $E$ with respect to $K$. After a long but  straightforward calculation the following equation is obtained
\begin{eqnarray}
\frac{H^{2}_{0}R_{0}}{\gamma}=\pi\Bigl[\sqrt{K^{2}-1}(\Bigl(-1+\frac{2(K^2-1)}{\mu-1}\Bigr)+K^{2}\arctan{\sqrt{K^{2}-1}}\Bigr]^{2}\\ \nonumber
\times\frac{\Bigl(2\sqrt{K^{2}-1}K(1+2K^{2})+(1-4K^{2})\log{\frac{K+\sqrt{K^{2}-1}}{K-\sqrt{K^{2}-1}}}\Bigr)}{K^{7/3}(K^{2}-1)^{2}(-3\sqrt{K^{2}-1}+(2+K^{2})\arctan{\sqrt{K^{2}-1}})}~.
\label{Eq:16}
\end{eqnarray}

Relation (2.21)  was derived in \citep{11} in a more compact form  using the demagnetizing field  coefficients of an ellipsoid and its adjoint.

\begin{figure}
\center
\includegraphics[width=0.8\textwidth]{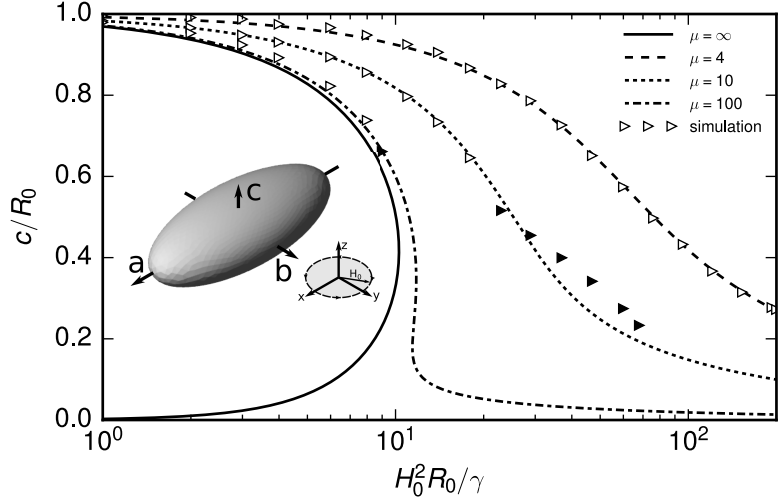}
\caption{Normalized short axis of the oblate shape ($c/R_{0}$) as a function of the magnetic Bond number for several values of $\mu=[4;10;100;\infty]$. Various lines are analytic  calculations for ellipsoidal  figures of equilibrium (obtained from section 2.3). Open triangles are the results of the numerical simulation of droplets with axial symmetry,  while solid triangles mark the  figures of equilibrium which are taking tri-axial shape (see section 4).}
\label{fig1}
\end{figure}

Figure~\ref{fig1} shows for several values of the magnetic  permeability the analytic dependence of the ratio $c/R_{0}=K^{-2/3}$\footnote {We have chosen to plot $c/R_0$ instead of $K$ since it remains a well defined quantity when the droplet takes a tri-axial shape.} on the magnetic Bond number $Bm=H^{2}_{0}R_{0}/\gamma$ ($R^{3}_{0}=abc$). It shows  that if the magnetic permeability is large enough, there is a hysteresis of $c/R_{0}$ as a function of the field. At infinite magnetic permeability an equilibrium shape, in the form of an oblate ellipsoid of revolution, only exists for magnetic fields below a critical field value.

In this respect the behavior is similar to that of a conducting droplet in an electrical field. An   equilibrium shape, elongated along the field, only exists if the electrical field strength is below a critical value \citep{7}. Above it there are jets appearing from the tips of the droplet.

Since in our case a fingering is observed experimentally (see figure~\ref{fig:Rot_ims} and \cite{8}) on the perimeter of the disc-like droplet  then it may be related to the threshold of existence of oblate equilibrium shape for large values of the 
magnetic permeability.

\begin{figure}
\center
\includegraphics{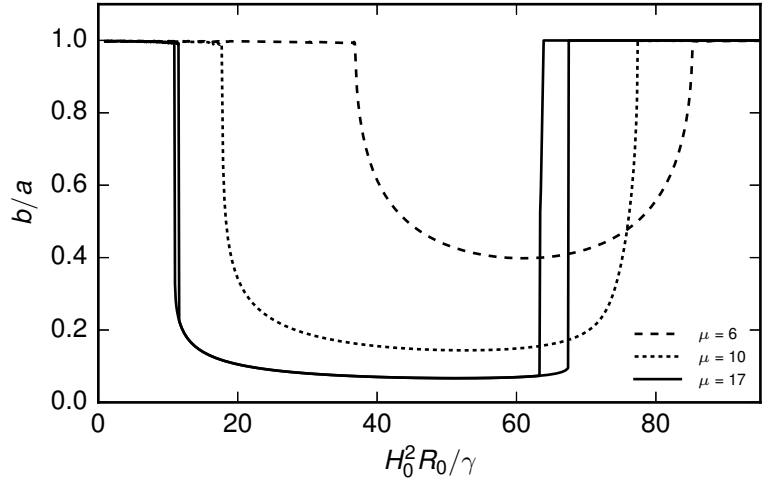}
\caption{Axis ratio $b/a$ of a three-axis ellipsoid as  a function  magnetic Bond number for several values of magnetic permeability  analytically calculated according to relation (2.22) with $\mu=17,10,6$. This illustrates the transitions from oblate ($a=b$) shape to tri-axial ellipsoid and then back to oblate shape, according to \citep{11}. It shows a hysteresis at large magnetic Bond numbers for $\mu>11$.} 
\label{fig2}
\end{figure}

By \cite{8} (see \cite{10} for more details) it was found during the analysis of the experimental results that the oblate shape loses its stability with respect to the symmetry destroying perturbations, the long axis of the ellipsoid of revolution remaining in the plane of the rotating field. The main conclusions included a critical value of the magnetic permeability for this instability and the re-entrant character of this oblate-prolate transition. This approach was further developed by \cite{11} where a general ellipsoid with the semi-axes ($a>b>c$)  for  symmetry destroying perturbations was considered. The main features of the oblate-prolate transition, apart from the prediction of hysteresis in the oblate-prolate and the prolate-oblate transitions, qualitatively remained unchanged. Here we are using these results (see figure \ref{fig2}) for the comparison with the numerical simulation of the three-dimensional droplet and experiments (see ahead in figure \ref{fig:three}) . For completeness we give the main points of the analysis following to \cite{11} approach. The energy of the droplet reads 
\begin{equation}
\frac{E}{2\pi\gamma R^{2}_{0}}=\Bigl(\frac{c^{2}}{ab}\Bigr)^{2/3}\Bigl(1+\frac{b}{\sqrt{a^{2}-c^{2}}}\Bigl(\rm{E}(k,m)\Bigl(\frac{a^{2}}{c^{2}}-1\Bigr)+\rm{F}(k,m)\Bigr)\Bigr)-\frac{1}{6}Bm\chi\Bigl(\frac{1}{1+4\pi\chi n_{x}}+\frac{1}{1+4\pi\chi n_{y}}\Bigr)~,
\label{Eq:17}
\end{equation}
where $k=\arcsin{\sqrt{1-c^{2}/a^{2}}};m=(1-(c/b)^{2})/(1-(c/a)^{2})$ and $\rm{E},\rm{F}$ are the elliptic integrals. We are using relations \citep{16} that express the demagnetizing field coefficients $n_{x},n_{y}$ in terms of the demagnetizing field coefficient $n_{z}$:
\begin{equation}
n_{z}(c/a,c/b)=\frac{1}{2}\int^{\infty}_{0}\frac{du}{(u+1)^{3/2}((c^{2}u/a^{2}+1)(c^{2}u/b^{2}+1))^{1/2}}~;
\label{Eq:18}
\end{equation}
\begin{equation}
n_{x}(c/a,c/b)=n_{z}(a/c,a/b);~
n_{y}(c/a,c/b)=n_{z}(b/a,b/c)~.
\label{Eq:18a}
\end{equation}
The equilibrium shapes are found by minimizing the energy of the droplet
with respect to $c/a$ and $c/b$. The bifurcation diagram in coordinates $H^{2}_{0}R_{0}/\gamma$ and $b/a$ for several $\mu$ values ($\mu=6,10,17$) is given in figure~\ref{fig2}. A hysteresis in the oblate-prolate-oblate  transition at large values of the magnetic field strength emerges for large values of $\mu$ which is in accord with the results obtained by \cite{11}. For small field values the hysteresis is small. The calculations are in agreement with \citep{11} and  show that the transition from oblate to a triaxial ellipsoid exists above some critical value of the magnetic susceptibility (in the interval of $\mu$ $[5;6]$). 

\section{\label{sec:num}Numerical algorithm}

\subsection{\label{sec:bound}Boundary integral equations (BIE) for magnetic field}

Since $\nabla\times\vec{H}=0$, we recall that the magnetic field is potential $\vec{H}=\nabla\psi$ and the potential due to $\nabla\bcdot\vec{H}=0$ satisfies the Laplace equation $\Delta\psi=0$. This equation can also be written in a boundary integral form \citep{19}
\begin{equation}
\int_{\partial \Omega} \frac{\partial \psi}{\partial n_{\boldsymbol x}} \frac{1}{|\boldsymbol x - \boldsymbol y|} dS_{\vec{x}}
-\int_{\partial \Omega} \psi(\boldsymbol x) \frac{\partial}{\partial n_{\boldsymbol x}} \frac{1}{|\boldsymbol x - \boldsymbol y|} dS_{\vec{x}}
=\left\{
    \begin{array}{ll}
      4\pi \psi(\boldsymbol y) ~&if~\vec{y} \in \Omega \\
      2 \pi \psi(\boldsymbol y)~&if~\vec{y} \in \partial \Omega \\
      0 ~&if~\vec{y} \ni \Omega 
    \end{array}
  \right.
\label{Eq:green}
\end{equation}
where $\Omega$ is an arbitrary region where $\Delta\psi=0$ and $\partial \Omega$ denotes the boundary of the fluid  drop. If $\vec y \in \partial \Omega$ the second integral from the left side must be interpreted in the Cauchy principal value sense which we denote here as $\dashint$. 

Applying the boundary conditions $\psi^{(i)}=\psi^{(e)}$ and $\mu\frac{\partial\psi^{(i)}}{\partial n}=\frac{\partial\psi^{(e)}}{\partial n}$ for the boundary of fluid drop $\partial \Omega$ we obtain the following boundary integral equation (BIE) for the magnetic potential (see appendix A)
\begin{equation}
\psi(\vec{y})=\frac{2\vec{H}_{0}\bcdot\vec{y}}{\mu+1}-\frac{1}{2\pi}\frac{\mu-1}{\mu+1}\dashint_{\partial\Omega}\psi(\vec{x})\frac{\partial}{\partial n_{\vec{x}}}\frac{1}{|\vec{x}-\vec{y}|}dS_{\vec{x}}~,
\label{Eq:30}
\end{equation}
where $\vec y \in \partial \Omega$. After solving this equation for $\psi$ we can use numerical differentiation to obtain the tangential components of the magnetic field $(\boldsymbol I - \boldsymbol n \otimes \boldsymbol n)\boldsymbol H$.

Numerous boundary integral equations exist for the normal component of the magnetic field. One possibility is to use \eqref{Eq:green} for the normal component of the field with a known $\psi$. However this gives an integral equation of the first kind and numerical experiments show a high sensitivity to the precision of the $\psi$ values. Another way is to formulate the boundary integral equation for the normal field as previously done for fluid droplets suspended in an electric field by \citet{BAYGENTS} (see also appendix \S~\ref{sec:bound-integr-equat} for derivation). However, it requires the integration of a strong singularity which is hard to do numerically in the general 3D case. Eventually we settled on interpreting a perturbation of the magnetic field from the droplet as coming from the surface currents on its interface $4 \pi \vec{K}/c = \vec n \times (\vec B^{(e)} - \vec B^{(i)}) = -(\mu - 1)\vec n \times \vec{H}$.

The magnetic field outside the fluid droplet is given by the Biot-Savart integral $\vec B = \vec B_0 + c^{-1}\int_{\partial \Omega} \vec K \times \nabla_{\vec{x}} \frac{1}{|\vec x - \vec y|} dS_{\vec x}$, which for the  normal component of the magnetic field in the presence of an external field $\boldsymbol H_0$ gives
\begin{eqnarray}
  \vec B \vec \bcdot \vec n_{\vec y} = \vec B_0 \vec \bcdot \vec n_{\vec y} + \vec n_{\vec y} \vec \bcdot \int_{\partial \Omega} \vec{K} \times \nabla_{\vec x}\frac{1}{|\vec x - \vec y |} dS_{\vec x} 
  = \vec B_0 \vec \bcdot \vec n_{\vec y} \\ \nonumber 
-\frac{\mu - 1}{4 \pi} \vec n_{\vec y} \vec \bcdot \int_{\partial \Omega} (\vec n_{\vec x} \times \vec H) \times \nabla_{\vec x}\frac{1}{|\vec x - \vec y |} dS_{\vec x}~,
\label{Eq:50}
\end{eqnarray}
which we also derive in appendix \S~\ref{sec:bound-integr-equat} without the use of the surface current interpretation. Since the integral on the right side of relation (3.3) only has a cross product of the field and the  normal, it allows us to calculate the normal field component from the known tangential components. The numerical algorithm for this calculation is described in \S~\ref{sec:num1}.

\subsection{\label{sec:num1} Numerical implementation}

Solving the BIE for the magnetic potential in the form (\ref{Eq:30}) or integrating (3.3) would require singular integral quadrature which makes these equations hard to implement numerically. We overcome this difficulty using a singularity subtraction technique, which is a common practise in the application of BIE \citep{18}. Specifically, for the magnetic potential we have an identity $\dashint \frac{\partial}{\partial n_{\vec x}} \frac{1}{|\vec x - \vec y|}dS_{\vec x} = - 2 \pi$ (see appendix \S~\ref{sec:sing-subtr}) which we use to rewrite \eqref{Eq:30} in the following form
\begin{equation}
 \psi(\boldsymbol y) \left( 1 -\frac{\mu -1 }{\mu + 1} \right) = \frac{2 \boldsymbol H_0 \bcdot \boldsymbol y}{\mu + 1} - \frac{1}{2 \pi} \frac{\mu - 1}{\mu + 1} \int_{\partial \Omega} \left[
 \psi(\boldsymbol x) - \psi(\boldsymbol y)
\right] \frac{\partial}{\partial n_{\boldsymbol x}} \frac{1}{|\boldsymbol x - \boldsymbol y|} dS_{\boldsymbol x}~,
\label{eq:Potential-BIE-regular}
\end{equation}
where $\dashint \rightarrow \int$ since we reduced the singularity by one order. The discretization of the drop surface into plane triangles enables us to solve it for arbitrary geometries.

The simplest quadrature for triangles $\int_{\Delta} f dS = (f_1 + f_2 + f_3) S_{\Delta}/3$ (trapezoidal quadrature) is used and it allows us to rearrange the integration efficiently. Denoting by $S_j$ the $1/3$ of the  sum of the area of all triangles with a common vertex $j$ we can rewrite \eqref{eq:Potential-BIE-regular} in a discrete form
\begin{equation}
  \psi_j \left( 1 -\frac{\mu -1}{\mu + 1} \right) = \frac{2 \boldsymbol H_0 \bcdot \boldsymbol x_0}{\mu + 1} - \frac{1}{2 \pi} \frac{\mu - 1}{\mu + 1} \sum_{i \neq j} [\psi_i - \psi_j] \frac{(\boldsymbol x_j - \boldsymbol x_i)\bcdot \boldsymbol n_j }{|\boldsymbol x_i - \boldsymbol x_j|^3} S_j~,
\end{equation}
which we solve numerically with the LAPACK library. Using a triangulation of the ellipsoid (for which the normal vectors and the solution of the magnetostatic problem are known) we see that the solution is only moderately sensitive to the precision of the normal vectors. Therefore, we use a simple and robust method for calculating the vertex normal vectors \citep{Jin2005} which is
\begin{equation}
  \vec n_j \propto \sum_i \alpha_i \vec n_i~,
\end{equation}
where the summation is over the triangles which have a vertex $j$. In relation (3.6) $\vec n_i$ is the normal to the plane triangle and $\alpha_i$ is the angle of the triangle at the vertex $j$.

In order to obtain the tangential components of the magnetic field we use the numerical differentiation of the obtained potential. For each vertex $j$ we write a system of equations 
\begin{equation}
  (\nabla \psi)_{j} \bcdot (\boldsymbol x_i - \boldsymbol x_j) = \psi_i - \psi_j~,
\end{equation}
where $i$ are the neighbour vertices. This system is overdetermined for the three components of $\nabla \psi$ therefore the linear least squares method is applied. Tangential field components are then projected out of the estimated gradient $\rm{P}\vec{H}_{j} = (\boldsymbol I - \boldsymbol n_j \otimes \boldsymbol n_j) (\nabla \psi)_{j}$, where $P=(\boldsymbol I - \boldsymbol n_j \otimes \boldsymbol n_j)$ is the projection operator. The known tangential field components according to the relation (3.3) give us the normal component of the field as a Biot-Savart quadrature. 

Since the quadrature in (3.3) is strongly singular it is inefficient to do it numerically in this form. The general procedure for subtracting the singularity in the Biot-Savart integral was given by \citet{POZRIKIDIS2000}, but it requires  calculating the curvature by adding another issue of complexity. Instead of that we have developed a regularisation technique which does not require a curvature calculation.

In the appendix \S~\ref{sec:sing-subtr} we prove the identity
\begin{eqnarray}
  \dashint_{\partial \Omega} \nabla_{\vec{x}} \frac{1}{|\vec{x} - \vec{y}|}\times (\vec{n}_{\vec{x}} \times \vec{H}_{\vec{x}}) dS_{\vec{x}} =  \dashint_{\partial \Omega} (\rm{P} \vec{H}_{\vec{x}} - \rm{P} \vec{H}_{\vec{y}}) \times \left(
    \vec{n}_{\vec{x}} \times \nabla_{\vec{x}} \frac{1}{|\vec{x} - \vec{y}|} 
  \right) dS_{\vec{x}}
  \nonumber \\
  -
  \dashint_{\partial \Omega} ( \rm{P} \vec{H}_{\vec{x}} - \rm{P} \vec{H}_{\vec{y}}) \left(
    \vec{n}_{\vec{x}} \bcdot \nabla \frac{1}{|\vec{x} - \vec{y}|}
  \right) dS_{\vec{x}}  - 2 \pi \rm{P} \vec{H}_{\vec{y}}~,
\end{eqnarray}
where $\rm{P}H_{\vec{y}}$ is a constant vector. This identity enables us to evaluate the strongly singular quadrature on the left side of relation (3.8) with weakly singular quadratures on its right side. Using this formula for (3.3) gives us weakly singular form of Biot-Savart quadrature
 \begin{eqnarray}
  \vec B \vec \bcdot \vec n_{\vec y} = \boldsymbol H_0 \bcdot \boldsymbol n_{\boldsymbol y}
  - \frac{\mu - 1}{4 \pi} \vec n_{\vec y} \vec \bcdot \dashint_{\partial \Omega} (\rm{P} \vec{H}_{\vec{x}} - \rm{P} \vec{H}_{\vec{y}}) \times \left(
    \vec{n}_{\vec{x}} \times \nabla_{\vec{x}} \frac{1}{|\vec{x} - \vec{y}|} 
  \right) dS_{\vec{x}} \nonumber \\
  + \frac{\mu - 1}{4 \pi} \vec n_{\vec y} \vec \bcdot
    \dashint_{\partial \Omega} (\rm{P} \vec{H}_{\vec{x}} - \rm{P} \vec{H}_{\vec{y}}) \left(
    \vec{n}_{\vec{x}} \bcdot \nabla \frac{1}{|\vec{x} - \vec{y}|}
  \right) dS_{\vec{x}}~.
   \label{eq:Biot-Savart-integral-regular}
 \end{eqnarray}

 We integrate both integrals of \eqref{eq:Biot-Savart-integral-regular} with the trapezoidal rule for non-singular integrals and use a polar transformation for weakly singular integrals \citep{19}. Specifically, when the singularity is on $\vec x_1$ we have the quadrature formula
 \begin{eqnarray}
   & \int_{\Delta} \frac{q(\boldsymbol x)}{|\boldsymbol x - \boldsymbol x_1|} dS_{\boldsymbol x}
    =
   \frac{|(\boldsymbol x_2 - \boldsymbol x_1)\times (\boldsymbol x_3 - \boldsymbol x_1)|}{|\boldsymbol x_2 - \boldsymbol x_1|} \int_0^{\pi/2}\frac{\int_0^{1/(\cos \chi + \sin \chi)} q(\rho, \chi) d \rho}{\sqrt{\cos^2 \chi + B \sin 2 \chi + C \sin^2 \chi}} d \chi \nonumber \\
   & \boldsymbol x = \rho \cos \chi \boldsymbol x_1 + \rho \sin \chi \boldsymbol x_2 + (1 - \rho (\cos \chi + \sin \chi))\boldsymbol x_3
     \nonumber \\
   &q = \rho \cos \chi q_1 + \rho \sin \chi q_2 + (1 - \rho (\cos \chi + \sin \chi))q_3 \nonumber \\
   &B = (\boldsymbol x_2 - \boldsymbol x_1)\bcdot (\boldsymbol x_2 - \boldsymbol x_1)/|\boldsymbol x_2 - \boldsymbol x_1|^2; C = |\boldsymbol x_3 - \boldsymbol x_1|^2/|\boldsymbol x_2 - \boldsymbol x_1|^2~,
       \label{eq:3}
 \end{eqnarray}
 
where we use a Gauss-Legendre quadrature with $10$ points for the $\rho$ and $\chi$ variables.

Besides the magnetic field another important  quantity for the surface force calculation is the curvature as can be seen in \eqref{Eq:22}. However, it turns out that the explicit mean curvature calculation can be avoided since we are interested in an integral containing a curvature over a closed surface. In \citep{Pozrikidis2001}, \citep{Zinchenko1997} and appendix \S~\ref{sec:identity} the following identity is derived for the value of the integral at a surface point with the radius vector $\vec{y}$ ($\vec{r}=\vec{x}-\vec{y}$)
\begin{eqnarray}
  \int \left(
    \frac{1}{R_1} + \frac{1}{R_2}
  \right) \vec n_{\vec y} \vec \bcdot
  \left(
    \frac{\boldsymbol I}{r} + \frac{\boldsymbol r \otimes \boldsymbol r}{r^3}
  \right) \vec \bcdot \vec n_{\vec x} dS_{\vec x}
  =
  -\int \frac{\vec r \vec \bcdot (\vec n_{\vec x} + \vec n_{\vec y})}{r^3}\bcdot\\ \nonumber
   \left(
    1 - \frac{3 (\vec r \vec \bcdot \vec n_{\vec x}) (\vec r \vec \bcdot \vec n_{\vec y})}{r^2}
  \right) dS_{\vec x}~.
\end{eqnarray}
Putting this formula in \eqref{Eq:22}, \eqref{Eq:23} and using the identity $\int \left( \frac{\vec I}{r} + \frac{\vec r \otimes \vec r}{r^3} \right) \vec \bcdot \vec n_{\vec x} dS_{\vec x} = 0$ for the singularity subtraction we get the normal component of the velocity 
\begin{equation}
  \bar{f} = \frac{1 - 1/\mu}{8 \pi} B_n^2 + \frac{\mu-1}{8 \pi} H_t^2 ~;
\label{eq:4}
\end{equation}
\begin{eqnarray}
  \vec v \vec \bcdot \vec n_{\vec y}  = \frac{1}{8 \pi \eta} \int_{\partial \Omega} 
  \left(
    \frac{\vec n_{\vec x} \vec \bcdot \vec n_{\vec y}}{r} + \frac{(\vec r \vec \bcdot \vec n_{\vec x}) (\vec r \vec \bcdot \vec n_{\vec y})}{r^3}
\right)
  (\bar{f}_{\vec x} - \bar{f}_{\vec y}) dS_{\vec x}
  +\\\nonumber
   \frac{\gamma}{8 \pi \eta} \int_{\partial \Omega} \frac{\vec r \vec \bcdot (\vec n_{\vec x} + \vec n_{\vec y})}{r^3} \left(
    1 - \frac{3 (\vec r \vec \bcdot \vec n_{\vec x}) (\vec r \vec \bcdot \vec n_{\vec y})}{r^2}
  \right) dS_{\vec x}~.
  \label{eq:2}
\end{eqnarray}
Using the trapezoidal integration formula for all triangular elements for both integrals we obtain the normal component of the velocity which we use to proceed to the next time step.

\subsection{Averaged equations for high frequency}

As noted in section \S~\ref{sec:intro} we are interested in the behaviour of the droplets in a rotating field with a high frequency. In this limit  we can neglect the droplet shape variation during the field period. At equal droplet and surrounding fluid viscosities the numerical simulations have shown that the high frequency behaviour establishes at $\omega \gg 10 R_0 \eta/\gamma$.

To derive the corresponding high frequency equations let us set, as previously, the magnetic field rotation plane as the $xy$ plane
\begin{equation}
 \vec H_0 = H_0 \cos(\omega t) \vec e_{x} +H_{0} \sin(\omega t) \vec e_{y}~.
\end{equation}
Because of the superposition principle we can write a solution of the magnetostatic equations
\begin{equation}
  \vec H(\vec x, \vec H_0) = H_0 \cos(\omega t) \vec H_{x} + H_0 \sin(\omega t) \vec H_{y}~,
  \label{Eq:15j}
\end{equation}
where we set $\vec H_{x} = \vec H (\vec x,\vec e_{x})$ and $\vec H_{y} = \vec H (\vec x,\vec e_{y})$. With this solution we can now write the normal component of the instantaneous magnetic surface force \eqref{eq:4} as
\begin{equation}
  \bar f = \frac{1 - 1/\mu}{8 \pi} (H_0 \cos(\omega t) B_{xn} + H_0 \sin(\omega t) B_{yn})^2 + \frac{\mu -1}{8 \pi} (H_0 \cos(\omega t) \rm{P} \vec H_{x} + H_0 \sin(\omega t) \rm{P} \vec H_y)^2~.
\end{equation}
Averaging this force over a single period gives us
 \begin{equation}
   \langle \bar f \rangle = \frac{1 - 1/\mu}{16 \pi} H_0^2 (B_{xn}^{2} + B_{yn}^{2}) + \frac{\mu - 1}{16 \pi} H_0^2  (H_{xt}^{2} + H_{yt}^{2})~,
 \end{equation}
which can be inserted in (3.13) to obtain the average normal velocity in high frequency rotating field.

\subsection{Mesh maintenance and generation}

Within a small number of time steps, at which the surface markers were advanced, an uneven triangulation is produced. It makes the representation of the geometry unsatisfactory due to degenerate triangular elements affecting the precision of the force and velocity calculation. To overcome this difficulty mesh relaxation and stabilization methods have been used for viscous drop simulations in \cite{Cristini2001} and \cite{ZINCHENKO1999}. Here we use a general purpose mesh  stabilization/relaxation algorithm by  \cite{Brochu2009}, which is also available as an open source library named ElTopo.

This algorithm collapses edges which are smaller than $l_{min}$,  splits those which are larger than $l_{max}$ and flips  edges  if after an operation they become shorter. The algorithm also moves each vertex to its average of neighbour vertices while the movement of vertex is restricted to its estimated normal plane.

To ensure that these operations do preserve sharp features in places with a high curvature they are allowed only if the volume changes introduced by any operation are smaller than $\gamma_{vol}$. Also, to avoid  generation  of new degenerate triangles, an operation is not allowed if it introduces triangles  with an angle either smaller than $\alpha_{min}$ or larger than $\alpha_{max}$. 
In our studies we used the following set of mesh configuration  maintenance parameters 
\begin{eqnarray*} \label{eq1}
  l_{min} =  0.7 \xi;~    l_{max} = 1.5 \xi;~    \gamma_{vol} = 0.1 \xi^3;~  \alpha_{min} = 15^{0};~   \alpha_{max} = 120^{0} 
\end{eqnarray*}
where we set  $\xi$ as the average edge length of the initial mesh.

We used a simple Euler method for advancing markers to new positions and used a sufficiently small time step so that the volume changes are  less than $1 \%$ when the equilibrium is obtained. We observed that large time steps could also induce a random noise and that the main limitation for the step-size comes from the higher chance of getting a degenerate mesh in the middle of a simulation.

For starting the simulation we need an initial triangulation. We used the Distmesh library developed by \cite{Persson2004} for generating the initial triangulation from the analytical signed distance functions  taking only the average edge length $\xi$ as an argument. With this algorithm we successfully generated high quality meshes for spheres, ellipsoids and star-like shapes.

\section{\label{sec:res}Numerical simulation results}

\subsection{Worm-like equilibrium figures}

The compression of the droplet perpendicularly to the plane of the rotating field was used as a test for the developed  algorithm along with smaller tests for droplet extension in a constant field, relaxation dynamics and magnetic field calculation for the ellipsoid.

 The procedure for finding the figures of equilibrium is as follows. Starting with initial triangulation we let the simulation algorithm relax the shape until the following conditions are met
\begin{equation}
\frac{v_i}{v_{start}} < 0.01 ;~ \tau_i v_i < 0.01 R_0 ;~ \tau_i = \frac{\Delta t}{\log \frac{v_{i-1}}{v_i}}~,
\label{eq:13}
\end{equation}
where $\Delta t$ is the time-step, $v_i$ is the maximum velocity at the step $i$ and $v_{start}$ is the velocity when the relaxation of triangulation begins. First condition ensures that the velocities start to decay exponentially which is expected when the surface is close to an equilibrium. The second condition assuming that equilibrium is approached exponentially requires that the maximum  deviation from true equilibrium (when $t \to \infty$), for any point on the surface,  is smaller than $1/100$ of $R_0$.

For obtaining figures of equilibrium we used two methods. In the simplest one (ellipsoid relaxation) we started from an ellipsoid with semi-axes calculated by the energy \eqref{Eq:17} minimization. In the second more complex one (quasi-static simulation) we start from triangulation of a sphere and increase the magnetic field quasi-statically ($\Delta Bm=1$) letting the triangulation relax at each step until condition 
\eqref{eq:13} is met. 

\begin{figure}
\center
\includegraphics[width=1.0\textwidth]{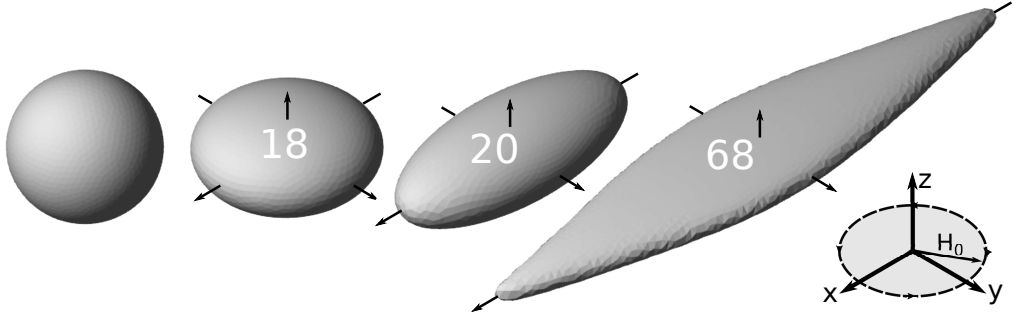}
\caption{Sequence of the established figures of equilibrium found numerically at different magnetic Bond numbers (0,18,20,68) with $\mu=10$.}
\label{fig5}
\end{figure}

We induced a spontaneous symmetry breaking of an axial figure, in a  quasi-static simulation, by stretching the triangulation in $x$ axis direction and compressing in the $y$ and $z$ axes directions to conserve the volume at the beginning of each quasi-step. When the difference between the lengths in the  $x$ and $y$ directions became larger than $0.01R_0$ we disabled this part of the algorithm. By disabling we kept triangulation more regular which, as we observed, slightly affects simulated equilibrium figures. We observed that more time steps are necessary for a quasi-step when the transition to a tri-axial figure is approached. The figures of equilibrium obtained with this algorithm at $\mu=10$ for different magnetic Bond numbers are shown in figure \ref{fig5}.

However, some differences are observed depending on how  the axial symmetry destroying perturbations are imposed - ellipsoid relaxation or quasi-static simulation.  Equilibrium shapes calculated by ellipsoid relaxation method are in excellent agreement with analytical results and also with experiments (as we will see in \S~ 5). The numerical experiments show that the difference between the figures of equilibrium obtained by ellipsoidal relaxation and by quasi-static simulation is due to an overly coarse triangulation and the quality of the mesh.

 \subsection{Star-like equilibrium figures}\label{star}

\begin{figure}
\center
\includegraphics[width=1.0\textwidth]{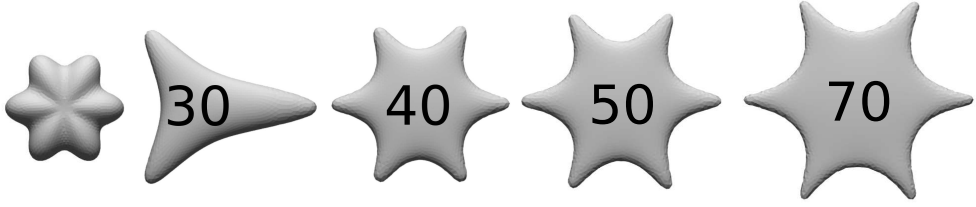}
\caption{Initial 6-star seed for simulations (left) and star-like droplet shapes obtained numerically at different magnetic Bond numbers with $\mu=10$.}
\label{fig8}
\end{figure}

Even if we do not manage to induce the star-like instability with numerical simulations directly from a worm-like figure, we are able to make a simulation of a star-like equilibrium figure, if we initially start with a mesh which has a broken symmetry at the starting point. As a seed we use the following surface
\begin{equation}
  x^2 + y^2 + 4 z^2 - \cos^2 \left( 3 \arctan \frac{y}{x} \right) \frac{x^2 + y^2}{x^2 + y^2 + z^2} =1~,
\label{eq4.2}
\end{equation}
which we meshed with the Distmesh library \citep{Persson2004}. We instantly set a magnetic field and depending on its strength we obtain the star-like equilibrium figures (see figure \ref{fig8}).

For small magnetic Bond numbers $Bm<19$ the perturbation decays and we obtain an axisymmetric oblate shape. At $Bm=30$ the 6-star perturbation still decays, but due to some imperfections in the initial mesh we obtain a triangular shape. Starting from magnetic Bond number $40$ we see that the 6-star perturbation grows for larger magnetic Bond numbers. Qualitatively the simulated star-like figures for different magnetic Bond numbers differ by their sizes in the $(x,y)$ plane but keep the length of the peaks approximately constant. 

With higher magnetic Bond numbers up to $100$ we were not able to show that the initial higher order perturbation decays. Instead some higher order perturbations determine the shape of the droplet as we expected from the experiment. However, we did calculate energies for the stars. Within error bars they agree with the quasi-static simulations for $Bm\leq 50$. However at $Bm=70$ we get an energy  $E/4 \pi R_0^2 \gamma=-3.15$ for the star, where for the worm-like shape we obtain $E/4 \pi R_0^2 \gamma=-2.98$.\footnote{ For numerical energy calculation we used formula (4.4). We tested this formula with ellipsoidal meshes. Comparing with analytical formula (2.19)  the error for the magnetic part was found to be about $2\%$ for a mesh with $3000$ triangles.}  Thus, the star-like shape has a smaller energy than the worm-like shape for magnetic Bond numbers $Bm\geq 70$. We can also conclude that by increasing the magnetic Bond number, the number of stable shapes increases while in the experiment we observe only two, when a hysteresis at the re-entrant transition happens.

\subsection{Numerical tests and results}

For the ratio $c/R_{0}$, which characterizes the compression of the droplet in a rotating field and is calculated according to relation (2.21), we see a qualitatively good agreement between the numerical results  and the analytic ellipsoidal approximation in a wide range of $\mu$ values as shown in figure \ref{fig1}. 
Specifically when $\mu=4$ the axially symmetric state of the droplet remains stable - the result which was predicted analytically for the droplet with an oblate ellipsoidal shape.

\begin{figure}
\center
\includegraphics{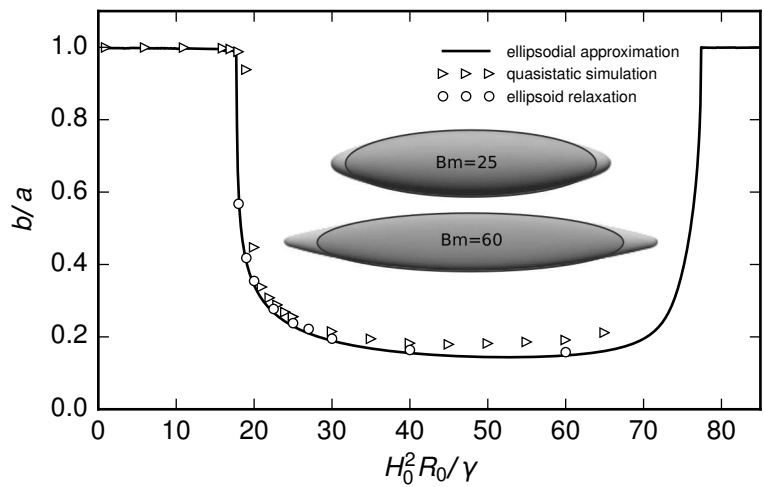}
\caption{Axis ratio of the droplet in the plane of the rotating field as a function of the magnetic Bond number with $\mu=10$. Full line: analytical  calculation using  ellipsoidal approximation (see section 2.3). Open circles- numerical data obtained  by ellipsoidal relaxation, triangles - numerical data obtained by quasi-static simulation. Equilibrium figures obtained by numerical simulation with  ellipsoid relaxation are shown in the figure. On top of them we have drawn ellipses to better see deviations at the tips.}
\label{fig6}
\end{figure}

On the other hand while the droplet keeps an ellipsoidal shape we may use the formula (2.22) for calculating the axis ratio in the plane of rotating field as can be seen in figure 6. We observe that the axis ratio $b/a$ starts to deviate from $1$ exactly where the ellipsoidal approximation predicts a transition from an oblate ellipsoid to a tri-axial one. We also see that both simulations very well agree with the analytic ellipsoidal approximation as expected since droplet in simulation keeps a quasi-ellipsoidal shape. For large magnetic Bond numbers the quasi-static simulation becomes unreliable because of mesh quality loss and in figure~\ref{fig6} we can see its growing difference with ellipsoid relaxation simulation.

Figure 6 also presents two calculated (ellipsoid relaxation) equilibrium figures with ellipses drawn on the top of them to show better their deviations. We can  see that figures of equilibrium starts to deviate significantly from the ellipsoidal shape for large magnetic fields, forming sharp tips as it is also observed in the experiments (figure 1). Unexpectedly the simulation closely follows the analytic ellipsoidal approximation even for large magnetic fields where the droplet presents tips. Thus we can use the ellipsoidal approximation quantitatively for adjusting the experimental data. 

We also calculate energies of the equilibrium figures and compare them with the energies found analytically in the ellipsoidal approximation \eqref{Eq:17}. This latter should be larger or equal due to constraints. The total instantaneous energy of the drop is the  sum of the surface energy $E_{S}=\gamma S$ and the magnetization energy $E_{M}$ \eqref{Eq:13}
\begin{equation}
  E = \gamma S  - \frac{1}{2} \int_{\Omega} \vec{M} \bcdot \vec{H}_0 dV = \gamma S - \frac{\mu - 1}{8 \pi} \int_{\Omega} \vec{H} \bcdot \vec{H}_0 dV = \gamma S - \frac{\mu - 1}{8 \pi} \int_{\partial \Omega} \psi \vec{H}_0 \bcdot \vec{n}_{\vec{x}} dS_{\vec{x}}~.
\end{equation}
Since we consider the high frequency limit of a rotating field, then according to the superposition principle (\ref{Eq:15j}), we write the average energy per period  in the following way
\begin{equation}
 \langle E \rangle = \gamma \int_{\partial \Omega} dS_{\vec{x}} - \frac{\mu-1}{8 \pi} \frac{H_0^2}{2} \int_{\partial \Omega} (\psi_x \vec{e}_x + \psi_y \vec{e}_y)\bcdot \vec{n}_{\vec{x}} dS_{\vec{x}}~,
  \label{eq:16}
\end{equation}
where $\psi_x(\vec{x})= \psi(\vec{e}_x,\vec{x})$ is the potential for a unit field directed along the $x$ axis. Numerical integration with the trapezoidal rule \eqref{eq:16} for each time-step showed that the energy decreases monotonously until it reaches its equilibrium value. 
\begin{figure}
\center
\includegraphics{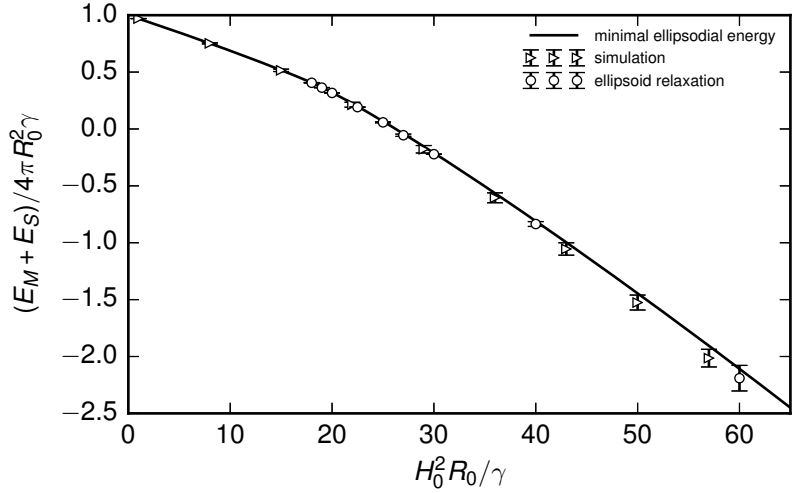}
\caption{Energy of droplets as a function of the magnetic Bond number. Comparison between numerical simulations with two different scenarios of inducing the symmetry destroying perturbations and analytical calculation for droplets of ellipsoidal shapes with $\mu=10$. Triangles are results of numerical simulation by inducing symmetry destroying perturbations as described in \S~4 and circle starting from triangulation of three axes ellipsoid with semi-axes calculated by the energy minimization. Solid line is deduced from the minimization of the energy given by (2.22). Difference is due to the non-uniformity of mesh in the first case.}
\label{fig4}
\end{figure}
We plot equilibrium energies in figure \ref{fig4} where we see that the energies of equilibrium shapes obtained in the quasi-static simulation and in the ellipsoidal approximation  \eqref{Eq:17} are within errorbars equal to the analytically calculated energy. The main contribution to the error comes from its magnetic part. For mesh with $3000$ triangles it equals to $2~\%$.

\begin{figure}
\center
\includegraphics{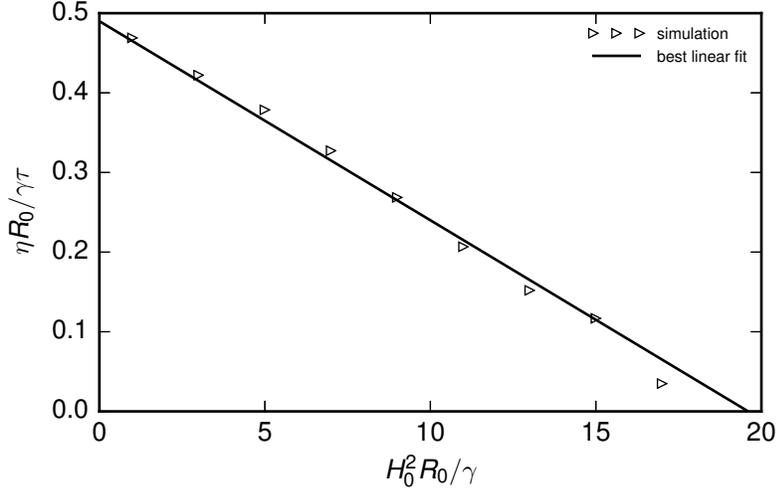}
\caption{Critical slowing down of relaxation of axial symmetry destroying perturbations in dependence on the magnetic Bond number. $\mu=10$. Triangles are results of numerical simulation.}
\label{fig3}
\end{figure}

As a last numerical result, we show here that the critical slowing down is a useful tool for the calculation of the critical magnetic field associated to the threshold of tri-axial ellipsoidal shape. By analogy with \citep{20} we can estimate the critical Bond number by calculating the decrement of the symmetry-breaking perturbations going to zero due to the critical slowing down at the bifurcation value of the magnetic Bond number. Using a simulation with $\mu=10$, which is close to the value for droplets used in experiments, we estimate the inverse of the  logarithmic decrement $\tau$ just before we jump to the next quasi-step. This decrement is drawn in figure \ref{fig3} in a normalized form. We clearly see that the drop shape with an axial symmetry becomes unstable around $Bm=19..20$ agreeing with result from ellipsoidal approximation and simulation (figure 6).

\section{\label{sec:exp}Experiments and their comparison with numerical simulation}
\subsection{Preparation of magnetic droplets}
Magnetic droplets can be prepared in various ways. The most common method is to disperse a magnetic fluid in an immiscible liquid \citep{26}. Recently an  interesting system was used where drops in glycerin were formed by magnetic nanoparticles with a biocompatible polydimethylsiloxane (PDMS) coating without a carrying liquid \citep{15}. However, here we use
magnetic drops that are obtained by inducing a phase separation in a magnetic fluid via an increase of ionic strength.
The water based magnetic fluid is prepared in the PHENIX laboratory, using the precipitation method \citep{Massart}, and has maghemite nanoparticles with a size $d=7.1$~nm and polydispersity $\sigma_{\rm PDI}=0.32$, volume fraction $\Phi=2.8$~\%, volume susceptibility $\chi=0.016$ (CGS units) and saturation magnetization $M_{\rm sat}=8.4$~G at $H_{sat}=10$~kOe, as given by magnetization measurements. 
The colloidal particles are stabilized with citrate ions from trisodium citrate, which is added in excess, creating a large initial ionic strength $I_{\rm citr}=0.18$~M present in the magnetic fluid.
By adding $0.05$~M of sodium chloride a phase separation is induced and droplets of concentrated phase are formed in the dilute phase.
As the droplets have a larger concentration of magnetic nanoparticles than the surrounding liquid in coexistence and a low surface tension, droplets respond to an external magnetic field of low amplitude.
A volume of $\approx 25$~$\mu$l of the phase separated fluid is put in quartz cuvettes of $0.1$~mm thickness (\textit{Helma analytics}). The sample is then investigated using the experimental setup.
\subsection{Experimental setup}
The experimental setup is based on a \textit{ZEISS} Axio Observer.D1 microscope equipped with a custom-built coil system.
Coils are powered by two \textit{KEPCO} BOP 20-10ML power supplies and controlled by \textit{NI} DAQ system.
The system has been calibrated with a gauss-meter, to link the magnetic field with current measurements. 
Images are recorded with \textit{AVT} Guppy F-046B (8-bit grayscale, resolution $640\times480$~px, framerate $3.75$~Hz).
All devices are governed by a custom-built program in \textit{NI} LabVIEW.
The system enables us to apply constant and alternating fields in the image plane.
A more detailed description of the experimental system and methods can be found in \citet{KitenbergsThesis}.

Recorded images are processed with MATLAB\textregistered.
They are first thresholded to binary images for detecting individual objects.
Then, to remove noise, dirt and other small drops, the largest object is selected and its perimeter is traced.
The longest distance between two points of the perimeter defines $2a$, while $2b$ is measured perpendicular to the midpoint of $2a$.

\subsection{Properties of magnetic droplets in constant field}
\label{sec:prop-magn-dropl}
\begin{figure}
\center
\includegraphics{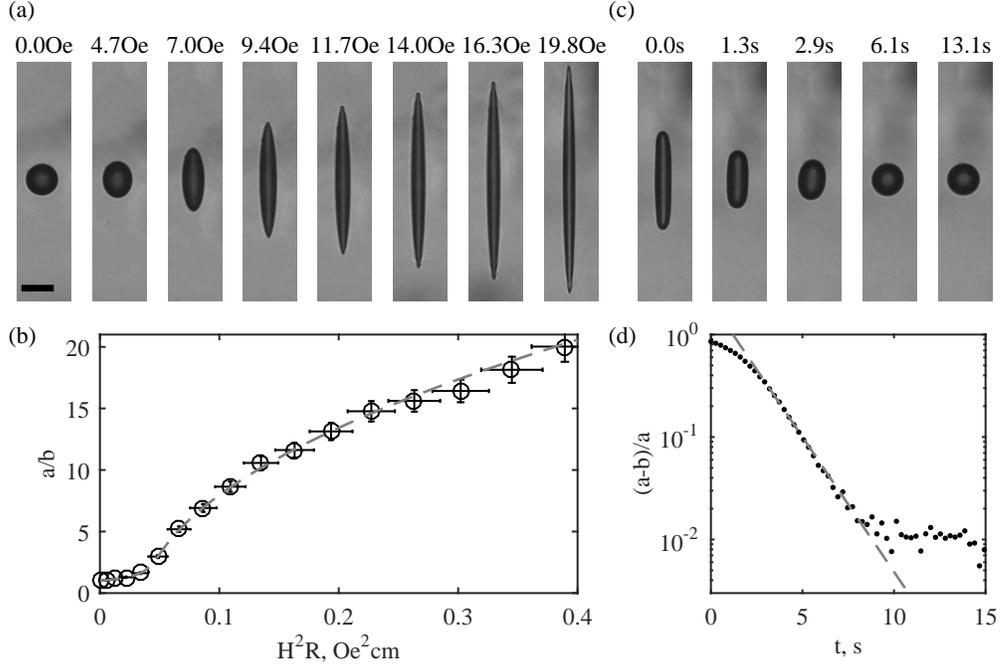}
\caption{An elongation-relaxation experiment with a typical droplet of initial diameter $R=10~\mu m$ and its analysis to obtain the properties of the drop (see text for details).Top: Pictures of the evolution (a) of the drop elongation under stepwise field increasing and of the temporal relaxation (c) of the droplet in zero field - scale bar is $20$~$\mu$m. Bottom: Evolution (b) of the droplet aspect ratio $a/b$ as a function of reduced field $H^{2}R$ and time relaxation (d) of the ratio $(a-b)/a$ in zero field. Dashed lines are adjustments by equations (5.1),(5.2) and (5.3).}
\label{fig:elong-relax}
\end{figure}

One way to experimentally obtain the physical properties of a magnetic drop, namely surface tension $\gamma$, magnetic permeability $\mu$ and viscosity $\eta$, is to make an elongation-relaxation experiment with subsequent analysis, as shown in figure~\ref{fig:elong-relax}.
For the elongation part, the magnetic drop is gradually deformed under a stepwise increase of a homogeneous and static magnetic field, with its direction in the image plane (figure~\ref{fig:elong-relax}~(a), parallel to the long side of the image).
After each step an image and current measurements are recorded once the steady state has been reached.
Image and data processing allows to form a plot of the ratio $a/b$ as a function of magnetic field, as shown in figure~\ref{fig:elong-relax}~(b).
The data are then adjusted with the equation from \citet{6} and \citet{eq431}, assuming an ellipsoidal shape of revolution and a constant volume for the droplet:
\begin{equation}
\displaystyle{H^2R=\gamma\left[\frac{4\pi}{\mu-1}+N\right]^2\frac{1}{2\pi}\frac{\left(\frac{3-2e^2}{e^2}-\frac{\left(3-4e^2\right)\arcsin{e}}{e^3\left(1-e^2\right)^\frac{1}{2}}\right)}{\left(1-e^2\right)^\frac{2}{3}\left(\frac{\left(3-e^2\right)}{e^5}\ln\left(\frac{1+e}{1-e}\right)-\frac{6}{e^4}\right)}},
\label{eq:elong}
\end{equation}
where $e$ is the ellipse eccentricity $(e=\sqrt{1-\frac{b^2}{a^2}})$ and $N$ its demagnetizing coefficient, which can be expressed for a prolate ellipsoid of revolution in the following way
\begin{equation}
N=\frac{4\pi\left(1-e^2\right)}{2e^3}\left(\ln\frac{1+e}{1-e}-2e\right).
\label{eq:demagnetizing}
\end{equation}
Equation \eqref{eq:elong}  allows us to obtain the $\gamma$ and $\mu$ values.

For the relaxation part, the magnetic field is cut to zero after a sufficient elongation and the droplet is allowed to relax back  to a spherical shape. Its relaxation is recorded over time (figure~\ref{fig:elong-relax}~(c)).
Close to a spherical shape, the parameter $(a-b)/a$ follows an exponential decay with a characteristic time $\tau_{\eta}$ as illustrated by figure~\ref{fig:elong-relax}~(d). In these experiments $(a-b)/a<0.02$ is the resolution limit, $\tau_{\eta}$ is described with an equation from \citet{drop-viscosity}:
\begin{equation}
\tau_{\eta}=\frac{R_0\left(16\eta_{d}+19\eta_c\right)\left(3\eta_{d}+2\eta_c\right)}{40\gamma\left(\eta_{d}+\eta_c\right)},
\label{eq:viscosity}
\end{equation}
where $R_0$ is the initial drop radius, $\eta_c$ is the viscosity of the drop (concentrated phase) and $\eta_d$ is the viscosity of the surrounding liquid (diluted phase).
$\eta_c$ can be found by taking $\eta_d$ as the viscosity of water ($\eta_d=0.01$~P) and using the surface tension $\gamma$ from the elongation part.
The phase-separated magnetic fluid droplets used in these experiments have a radius $(R_0\pm\Delta R_0)=(10.0\pm0.5)$~$\mu$m and properties $(\gamma\pm\Delta\gamma)=(6.0\pm 0.5)\bcdot10^{-3}~\rm{erg}/\rm{cm}^{2}$, $(\mu\pm\Delta\mu)=(10.5\pm 0.5)$ and $(\eta_c\pm\Delta\eta_c)=(10\pm 1)$~P. However, let us note that according to  \citep{21} equation (5.1) is only valid up to $a/b\approx 7$, thus hereafter we use calculated $\gamma$ deduced from the compression of the oblate shape.

\subsection{Properties of magnetic droplets in rotating field}

Magnetic drop deformation under a rotating field is also studied here, using a frequency $500$~Hz.
It is notably larger than the frequency of the droplet breathing mode $\gamma/R_0\eta_c\approx 0.5$~Hz fulfilling the high frequency criterion. Magnetic field strength is increased and decreased stepwise, allowing to reach equilibrium after each step. Equilibrium shape is recorded as a function of the field, together with the characteristic time to reach equilibrium. 

\begin{figure}
\center
\includegraphics{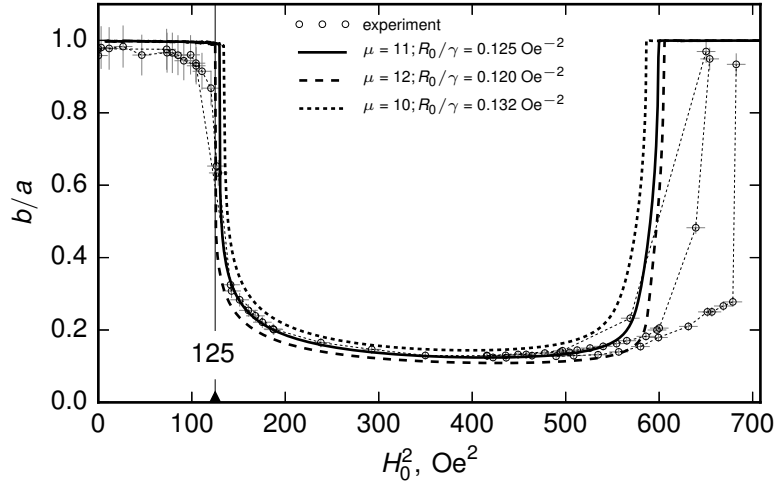}
\caption{Axes ratio 
$b/a$ of equilibrium droplet shapes, in the plane of rotating field,  as a function of the square of the field strength. Experimental data (circles) are compared with analytic calculations (black lines) using  ellipsoidal approximation formula (2.22). The calculation with $\mu=11$ and $R_0/\gamma=0.125~\rm{Oe}^{-2}$ fits the best. Thin dashed lines indicate separate experiments. The critical field value, which is determined by relaxation of symmetry destroying parturbations (see figure 11(b)) is marked with a triangle at $H_{0}^{2}=125~\rm{Oe}^{2}$. }
\label{fig:three}
\end{figure}

Both in numerical simulations and experiments we see that the oblate magnetic droplet becomes unstable with a respect to axial symmetry breaking perturbations at the critical magnetic Bond number. Above the threshold value of the magnetic Bond number it assumes a worm-like shape for small fields and a star-like shape for larger magnetic Bond numbers, possibly by a bifurcation of a different kind (see figure~\ref{fig:Rot_ims}). The  droplets axes ratio $b/a$ obtained experimentally in the range of moderate field strength is shown in figure~\ref{fig:three}. Separate experiments, indicated with thin dashed lines, show a good agreement up to a field $H^2\approx 550$~Oe$^2$. Above that, the droplet shape can remain elongated, but relaxes to the energetically more preferable star shape (as shown in section~\ref{star}), if the deformation via the increase of the field is large enough.

\begin{figure}
\center
\includegraphics{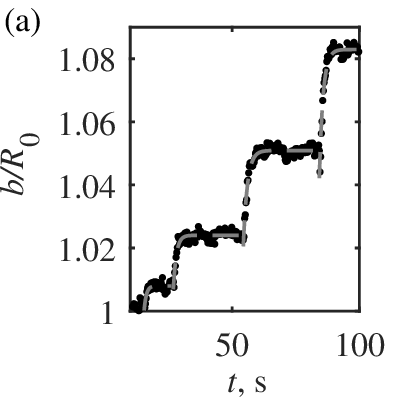}
\includegraphics{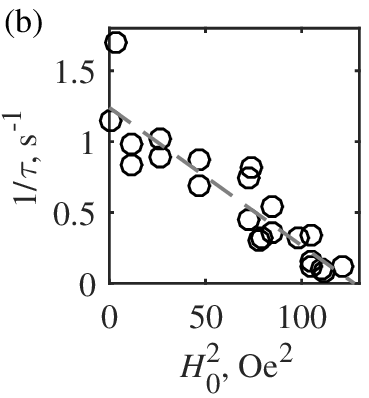}
\includegraphics{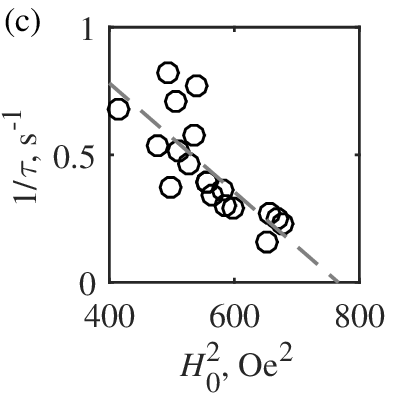}
\caption{(a) Shape perturbations resulting from small stepwise field increases. For each step an exponential decay with a characteristic time $\tau$ is observed. (b,c) Magnetic field dependence of the  characteristic decay for symmetry destroying perturbations - (b) approaching theoblate-prolate transition and (c) approaching the prolate-oblate transition. Dashed lines are best linear fits.}
\label{fig:crit_slow}
\end{figure}

The magnetic field is increased by small steps. Nevertheless, as shown in figure~\ref{fig:crit_slow}(a) the shape perturbations and their dynamics (black dots) can be visually detected, which allows to trace their characteristic decay time $\tau$ (dashed line). 
As in the numerical simulation (see figure~\ref{fig3}) the critical slowing down for the decay of perturbations in the case of an oblate shape is shown in figure \ref{fig:crit_slow}(b). It leads to a  critical field is $H_c^2=125~\rm{Oe}^2$. 
Experimentally it is possible to observe the critical slowing down of the prolate shape as shown in figure~\ref{fig:crit_slow}(c), when approaching prolate-oblate transition, where the critical field is $H_{c,2}^2=765~\rm{Oe}^2$.

The compression of the oblate shape is determined from the semi-axes of the visible projection of the droplet approximated by a 3D ellipsoid according to $c/R_{0}=R^{2}_{0}/ab$. By numerical experimentation with formula (2.21) we have found that semi-axes ratio $1/K$ is a linear function of $H_0^2$ in a broader field range than $c/R_0=K^{-2/3}$. Thus from experiment we deduce  $1/K = (ab/R_0^2)^{-3/2}$. In figure 12 $1/K$ is plotted as a function of $H^{2}_{0}$. It is adjusted with the relation 
\begin{equation}
\frac{1}{K}=1-\frac{H^{2}_{0}}{H^{2}_{*}},
\label{fig:fit}
\end{equation}
and $H^2_{*}=303~\rm{Oe}^2$. While according to (2.21) $H^{2}_{*}$ reads
  \begin{equation}
    H^{2}_{*}=\frac{64\pi}{9}\Bigl(\frac{\mu+2}{\mu-1}\Bigr)^{2}\frac{\gamma}{R_{0}}.
    \label{eq:14}
    \end{equation}

\begin{figure}
\center
\includegraphics{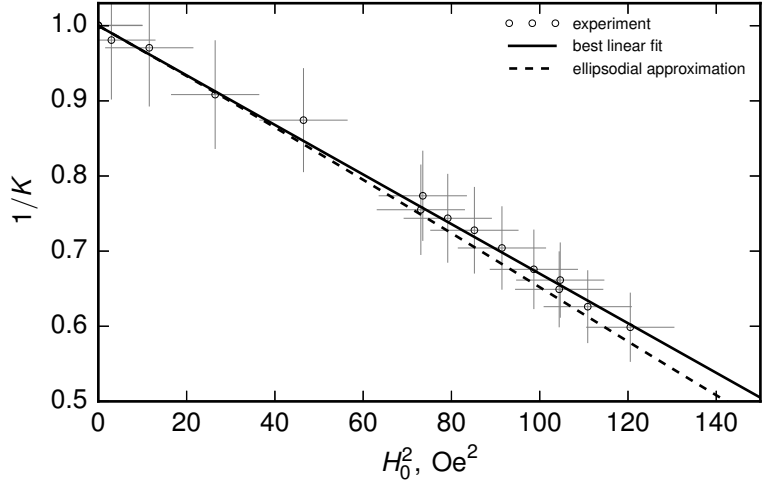}
\caption{Compression of the oblate shape in dependence on the square of field strength; Experimental data (dots) calculated with formula $1/K=(ab/R_0^2)^{-3/2}$; best linear fit of the data (solid line); Analytical calculation with ellipsoidal approximation (dashed line) with $\mu=11$ and $R_0/\gamma=0.125~\rm{Oe}^{-2}$.}
\label{fig:exp}
\end{figure}

In section 4 it is shown that the ellipsoidal approximation for the calculation of the axis ratio $b/a$ is valid even when the droplet becomes non-ellipsoidal. Thus we use the ellipsoid approximation to fit the data, adjusting $\mu$ with constrained $R_0/\gamma$ by relation \eqref{eq:14}. In figure 10 we plot the ratio $b/a$ obtained analytically from the ellipsoidal approximation for $\mu=10,11,12$. We see that theory gives a very good agreement when $\mu=11$. This value also agrees with the elongation experiment we did in \S~\ref{sec:prop-magn-dropl}. 

When we take $\mu=11$ and $R_0=10~\mu\rm{m}$, we get the surface tension from (\ref{eq:14})  $\gamma = 8.0 \cdot 10^{-3}~\rm{erg}/\rm{cm}^2$ ($R_0/\gamma=0.125~\rm{Oe}^{-2}$ in reduced numerical unit). This value is slightly larger than the one calculated from the  elongation experiment. But as we see in figure~\ref{fig:three} that it fits better the axes ratio data. We take this value to evaluate magnetic Bond number in the rotating field experiments. 

Another peculiar aspect, for which we can use the results obtained thus far, is for the calculation of the critical magnetic Bond number $Bm_c$. According to figure 11 we have $H_c^2=125~\rm{Oe}^2$. It results in $Bm_c=15.6$ ($R_0/\gamma=0.125~\rm{Oe}^{-2}$), which is close to the value obtained numerically $Bm_c=16.2$. Since energy differences between axial shape and tri-axial are very small near the critical magnetic Bond number then the deviation can either come from numerical errors or some small physical effects in experiments.

As the typical properties of the droplets are now determined  we use in the following the magnetic Bond number $Bm$ to characterize the state of the droplet. The rich family of characteristic drop shapes observed during these experimental measurements is displayed in the introduction of this work (see figure~\ref{fig:Rot_ims}).

\subsection{Comparison of experimental and numerical results}

\begin{figure}
\center
\includegraphics[width=1.0\textwidth]{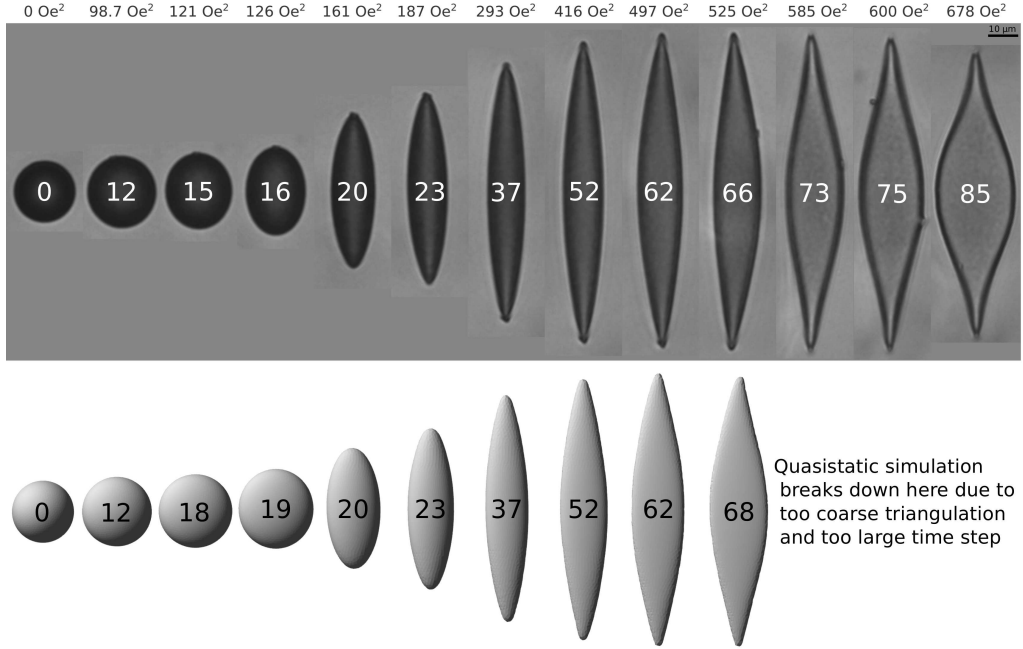}
\caption{ Qualitative comparison of equilibrium figures obtained numerically (bottom) and experimentally (top). Experiments correspond to $\mu=11$ and $R_0/\gamma = 0.125~\rm{Oe}^{-2}$. Numerical simulation is made with $\mu=10$. Black and white numbers on figures are magnetic Bond numbers. }
\label{fig7}
\end{figure}

We first compare qualitatively the equilibrium figures obtained experimentally   with the numerical simulation results from section 4  in figure~\ref{fig7}. Numerous effects can be seen in the experiments and in the simulations. (i) For magnetic Bond numbers under $Bm<20$ droplet becomes flatter  with increasing field strength as shown in figure~\ref{fig1} which is very well described with analytical results. Figure~ \ref{fig:exp} shows that this droplet flattening $1/K=c/a=f(H_{0})$ is almost a linear function of field strength square.
(ii) At $Bm_c=20$, there is a spontaneous symmetry breaking . (iii)  For $20<Bm<50$, the worm-like shape elongation increases with the field strength. (iiii)For $50<Bm<80$ a non-ellipsoidal regime  where the droplet becomes wider, flatter and forms sharp tips is observed. (iiiii) At $Bm=85$ a re-entrant transition is observed experimentally. We were not able to simulate it numerically since the  droplet becomes so flat that it would need more than $10000$ triangles to have a width  of at least three elements. It would require a very time consuming simulation. 

To explore the star-like shapes we started with an already broken symmetry mesh and put corresponding magnetic field on. We obtained shapes qualitatively similar to the experimental star-like shapes (see figure 5).  A more detailed investigation of hysteresis at the re-entrant transition is pending for a future publication. Here we only point out that its observation in numerical simulations is not easy due to the strong compression of the droplet at large magnetic Bond number values.

\begin{figure}
\center
\includegraphics[width=1.0\textwidth]{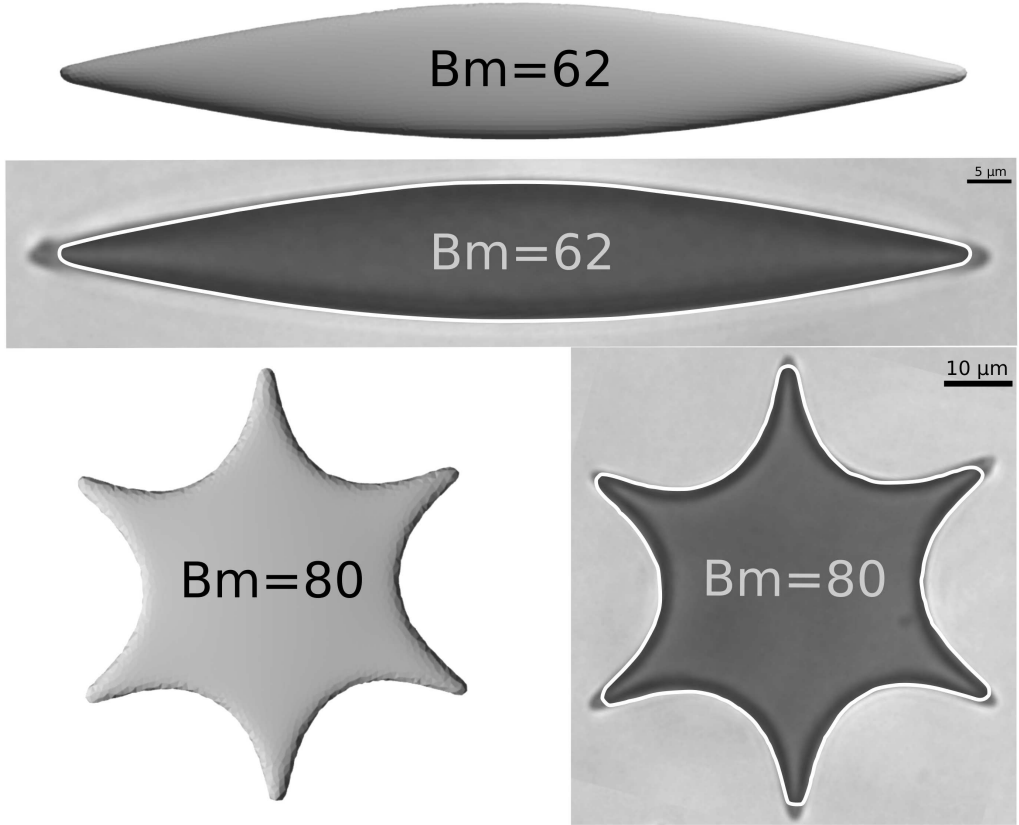}
\caption{Comparison of shapes observed in experiment and in numerical simulations.
The worm-like figure is obtained starting initially with an ellipsoid. The star is obtained starting from a seed, given by equation~\eqref{eq4.2}. To emphasize the differences and the similarities between experimental shapes and numerical ones, we plot as a white line the outline obtained from simulations on the two  experimental images. $\mu=11$ and $R_0/\gamma=0.125~\rm{Oe}^{-2}$. }
\label{fig:rel-exp}
\end{figure}

As an important result we present a quantitative comparison of  equilibrium figures obtained experimentally and numerically by drawing calculated equilibrium shapes  on top of experimental images  as shown in figure~\ref{fig:rel-exp}. Overall we see a good agreement except that simulation tips are a bit shorter either because of numerical errors or because $\mu$ is a bit larger in experiment.  

It is also worth noting the differences between experimental data and elliposidal approximation at larger fields, as visible in figure~\ref{fig:three} for $H^2>550$. For this we see several possible explanations. First, the drop shape at larger fields is no more ellipsoidal, but takes a rather flattened shape with sharper tips. Second, the surface tension of the concentrated magnetic particle system, might become anisotropic under large fields. Third, owing to different field increase steps, as visible in Fig.10, perturbations of different magnitude are imposed on the droplet changing the value  of field strength at which the re-entrant transition is observed.

\section{Conclusions and discussion}
As a general comment we would like to emphasize that the description of the highly magnetic droplets under study here, which are issued from a phase separation, is not obvious. They are in fact based on a new kind of soft magnetic matter with strong interparticle interaction. We show here  that many features of their under-field behaviour may be however described by a simple model of magnetic fluid.
Moreover it is shown here that the boundary integral equation technique allows one to develop efficient numerical algorithms for the simulation of the equilibrium shapes of these magnetic droplets submitted to an applied field either static  or rotating. Numerical and experimental results show a good agreement at equilibrium and as well during shape relaxation processes.  Under rotating field, many similar features are observed both in numerical simulations and experiments, namely - instability of the axially symmetric droplet shape with respect to the symmetry destroying perturbations (leading to the oblate-prolate transition), flattening of the prolate shape as the rotating  magnetic field strength is increased, formation of the star-fish like equilibrium shapes.

The experimental and numerical results show that the behaviour of the magnetic droplets may be well described by approximating their shape by a tri-axial ellipsoid.
However the observed deviations from the simple ellipsoidal approximation, such as the formation of sharp tips, challenge for their theoretical description. Besides that,
further development of this work is connected with accounting for the large viscosity contrast between the magnetic droplet and the surrounding fluid. It is possible by applying the boundary integral equation technique to the solution of Stokes equation. Another important issue which should be mentioned with a respect to these strongly magnetic droplets, issued from a phase separation, is the contribution of the magnetic dipolar interactions  to their surface tension. An estimate shows that this contribution is just of the same order of magnitude as the surface  tension determined in experiments and it may depend on the orientation of the interface with respect to the applied field.

\section*{Acknowledgements}
Authors are grateful to Delphine Talbot for providing us with the magnetic fluid sample and Rudolfs Livanovics for fruitful discussions. The research was funded by a project of  the University of Latvia "Nano and quantum  technologies, innovative materials - Soft materials in electromagnetic  field".

\appendix
\section{Boundary integral equations}\label{sec:bound-integr-equat}

\subsection{Boundary integral equation for potential}

For the potential boundary integral equation derivation we are going to use formula \eqref{Eq:green} which follows from the Green formula. With subscripts $\psi^{(i)}$ and $\psi^{(e)}$ we are denoting the limiting values of the potential approaching to the surface from both sides of the droplet boundary. For the limiting value from the inside we use relation \eqref{Eq:green} when $\vec{y}\in \partial \Omega$
\begin{equation}
  2 \pi \psi^{(i)}(\vec{y}) = \int_{\partial \Omega} \frac{\partial \psi^{(i)}}{\partial n_{\vec{x}}} \frac{1}{|\vec{x} - \vec{y}|} dS_{\vec{x}} - \dashint_{\partial \Omega} \psi^{(i)}(\vec{x}) \frac{\partial}{\partial n_{\vec{x}}} \frac{1}{|\vec{x} - \vec{y}|} dS_{\vec{x}}
  \label{eq:8}
\end{equation}
and for the external region $\Omega_e = \infty - \Omega$ when $\vec{y} \in \partial \Omega$
\begin{eqnarray}
  2 \pi \psi^{(e)}(\vec{y}) = - \int_{\partial \Omega} \frac{\partial \psi^{(e)}}{\partial n_{\vec{x}}} \frac{1}{|\vec{x} - \vec{y}|} dS_{\vec{x}} + \dashint_{\partial \Omega} \psi^{(e)} \frac{\partial}{\partial n_{\vec{x}}} \frac{1}{|\vec{x} - \vec{y}|} dS_{\vec{x}}  \nonumber \\
  +  \int_{\partial \infty} \frac{\partial \psi}{\partial n_{\vec{x}}} \frac{1}{|\vec{x} - \vec{y}|} dS_{\vec{x}} - \int_{\partial \infty} \psi \frac{\partial}{\partial n_{\vec{x}}} \frac{1}{|\vec{x} - \vec{y}|} dS_{\vec{x}}~,
  \label{eq:5}
\end{eqnarray}
where for the first two integrals we changed the direction of the normal vector so it is directed outwards of $\Omega$. The last two integrals of \eqref{eq:5} according to \eqref{Eq:green} correspond to the external field contribution to the potential. Since here we are interested in case of homogenous field whose potential is $\vec H_0 \bcdot \vec{y}$ then we rewrite \eqref{eq:5} 
\begin{eqnarray}
  2 \pi \psi^{(e)}(\vec{y}) = 4 \pi \vec H_0 \bcdot \vec{y} - \int_{\partial \Omega} \frac{\partial \psi^{(e)}}{\partial n_{\vec{x}}} \frac{1}{|\vec{x} - \vec{y}|} dS_{\vec{x}} + \dashint_{\partial \Omega} \psi^{(e)} \frac{\partial}{\partial n_{\vec{x}}} \frac{1}{|\vec{x} - \vec{y}|} dS_{\vec{x}}~.  
  \label{eq:9}
\end{eqnarray}
Multiplying equation \eqref{eq:8} by $\mu$ and adding it to \eqref{eq:9} and using the boundary conditions $\psi^{(i)} = \psi^{(e)}$ and $\frac{\partial \psi^{(e)}}{\partial n} = \mu \frac{\partial \psi^{(i)}}{\partial n}$ we obtain the boundary integral equation for the potential
\begin{equation}
  \psi(\vec{y})=\frac{2\vec{H}_{0}\bcdot\vec{y}}{\mu+1}-\frac{1}{2\pi}\frac{\mu-1}{\mu+1}\dashint_{\partial\Omega}\psi(\vec{x})\frac{\partial}{\partial n_{\vec{x}}}\frac{1}{|\vec{x}-\vec{y}|}dS_{\vec{x}}.
  \label{eq:11}
\end{equation}
where we omit the suffixes since the potential is continuous. 

\subsection{Boundary integral formula for normal component}

Looking for the solution of the Laplace equation in the form of a single layer potential it is possible to derive the boundary integral equation for the normal component of the magnetic field strength. Numerical calculations showed that the solution of this equation is ill-behaved and so we have chosen an alternative approach deriving an integral relation which allows us to calculate the normal component of the magnetic field strength from the known tangential component.

Let us consider the identity
\begin{equation}
\nabla_{\vec{x}}\bcdot (\vec{Q}\times\vec{B})=\vec{B}\bcdot \nabla_{\vec{x}}\times \vec{Q}~,
\label{Eq:31}
\end{equation}
where $\vec{B}$ is the magnetic induction, $\vec{Q}=\nabla_{\vec{x}}\times\frac{\vec{a}}{|\vec{x}-\vec{y}|}$ ($\vec{a}$ is an arbitrary constant vector. It is easy to check that $\nabla_{\vec{x}} \times \vec{Q}=(\vec{a}\bcdot \nabla_{\vec{x}})\nabla_{\vec{x}}\frac{1}{|\vec{x}-\vec{y}|}$. Integrating relation (\ref{Eq:31}) throughout the arbitrary region $\Omega$ in dependence on the position of the point with the radius vector $\vec{y}$ we have
\begin{equation}
\int_{\partial \Omega}(\vec{B} \times\vec{n}_{\vec{x}})\times\nabla_{\vec{x}}\frac{1}{|\vec{x}-\vec{y}|}dS_{\vec{x}}
-\int_{\partial \Omega}(\vec{B}\bcdot\vec{n}_{\vec{x}})\nabla_{\vec{x}}\frac{1}{|\vec{x}-\vec{y}|}dS_{\vec{x}}
=\left\{
    \begin{array}{ll}
      4\pi \vec{B}(\vec y)~&if~\vec{y} \in \Omega \\
      2 \pi \vec{B}(\vec y)~&if~\vec{y} \in \partial \Omega \\
      0 ~&if~\vec{y} \in \overline{\Omega \cup \partial \Omega}
    \end{array}
  \right.
\label{Eq:31a}
\end{equation}
where $\vec{n}$ is the external normal to the boundary of the region and the integrals in the case $\vec{y}\in\partial\Omega$ are taken as the Cauchy principal value integrals.

Applying relation (\ref{Eq:31a}) for the region $\Omega_{e}=\infty -\Omega$, where $\Omega$ is the region occupied by the magnetic droplet at $\vec{y} \in \partial\Omega$ we have (at $\partial\Omega$ we take $\vec{n}$ as the external normal to the surface of the droplet)
\begin{eqnarray}
  2\pi\vec{B}^{(e)}(\vec{y})=-\dashint_{\partial \Omega}[\vec{B}^{(e)}\times\vec{n}_{\vec{x}}]\times\nabla_{\vec{x}}\frac{1}{|\vec{x}-\vec{y}|}dS_{\vec{x}} + \dashint_{\partial \Omega}(\vec{B}^{(e)}\bcdot\vec{n}_{\vec{x}})\nabla_{\vec{x}}\frac{1}{|\vec{x}-\vec{y}|}dS_{\vec{x}} \nonumber \\ 
  +
\int_{\partial \infty}(\vec{B}\times\vec{n}_{\vec{x}})\times\nabla_{\vec{x}}\frac{1}{|\vec{x}-\vec{y}|}dS_{\vec{x}}
  -\int_{\partial \infty}(\vec{B}\bcdot\vec{n}_{\vec{x}})\nabla_{\vec{x}}\frac{1}{|\vec{x}-\vec{y}|}dS_{\vec{x}}~.
  \label{eq:10}
\end{eqnarray}
According to relation \eqref{Eq:31} last two integrals correspond to contribution of the external field. Since we have a homogeneous field $\vec{H}_0$ (without droplet) everywhere then we rewrite relation \eqref{eq:10} as 
\begin{equation}
2\pi\vec{B}^{(e)}=-\dashint_{\partial \Omega}(\vec{B}^{(e)}\times\vec{n}_{\vec{x}})\times\nabla_{\vec{x}}\frac{1}{|\vec{x}-\vec{y}|}dS_{\vec{x}}
+\dashint_{\partial \Omega}(\vec{B}^{(e)}\bcdot\vec{n}_{\vec{x}})\nabla_{\vec{x}}\frac{1}{|\vec{x}-\vec{y}|}dS_{\vec{x}}+4\pi\vec{H}_{0}~,
\label{Eq:31d}
\end{equation}
where for simplicity we can say that $\vec{H}_0$ corresponds also to magnetic field far away of droplet.

Taking the projection of relation \eqref{Eq:31a} at $\vec{y}\in\partial\Omega$ and \eqref{Eq:31d} on the normal to the interface and accounting for the boundary conditions $\vec{B}^{(e)}\bcdot\vec{n}=\vec{B}^{(i)}\bcdot\vec{n}$ and $\vec{H}^{(i)}\times\vec{n}=\vec{H}^{(e)}\times\vec{n}$ we obtain the following relation for the calculation of the normal component of the magnetic field
\begin{eqnarray}
  \vec B \vec \bcdot \vec n_{\vec y}  = \vec H_0 \vec \bcdot \vec n_{\vec y} 
- \frac{\mu - 1}{4 \pi} \vec n_{\vec y} \vec \bcdot \int_{\partial \Omega} (\vec n_{\vec x} \times \vec H) \times \nabla_{\vec x}\frac{1}{|\vec x - \vec y |} dS_{\vec x}~.
\label{eq:1}
\end{eqnarray}

\subsection{Boundary integral equation for normal component of field}
\label{sec:bound-integr-equat-2}

While we had a lack of success with direct calculation of the normal field with the boundary integral equation in our studies we have derived a corresponding equation for 3D. Here we briefly give the derivation as it is analogous and uses previous results of derivations \eqref{eq:11} and \eqref{eq:1}.

Again taking the normal projection of relation \eqref{Eq:31a} at $\vec{y} \in \partial \Omega$ and \eqref{Eq:31d} and using the boundary conditions $\vec{B}^{(e)}\bcdot\vec{n}=\vec{B}^{(i)}\bcdot\vec{n}$ and $\vec{H}^{(i)}\times\vec{n}=\vec{H}^{(e)}\times\vec{n}$ we subtract $\int_{\partial \Omega} (\vec B \times \vec{n}_{\vec{x}}) \times \nabla_{\vec{x}} \frac{1}{|\vec{x} - \vec{y}|} dS_{\vec{x}}$ and obtain
\begin{equation}
  B_n(\vec{y}) = \frac{2 \mu \vec{H}_0 \bcdot \vec{n}_{\vec{y}}}{\mu + 1} + \frac{1}{2 \pi} \frac{\mu - 1}{\mu + 1} \dashint_{\partial \Omega} B_n(\vec{x}) \frac{\partial}{\partial n_{ \vec{y}}} \frac{1}{|\vec{x} - \vec{y}|} dS_{\vec{x}} ~,
\end{equation}
where $B_n = \vec{B}\bcdot \vec{n}$. This equation is remarkably similar to the boundary integral equation for potential \eqref{eq:11} except for normal derivative inside the integral. Using the same numerical procedure as outlined in \S~\ref{sec:num} (plane triangles, trapezoidal integration for all triangles) we report that this integral equation gives large errors. This is clearly because we don't integrate the singular quadrature properly which is cumbersome for 3D. 

\section{Singularity subtraction}\label{sec:sing-subtr}

Consider identity with $\vec y \ni \Omega$
\begin{equation}
  \int_{\partial\Omega}\vec{n}_{\vec{x}} \bcdot \nabla_{\vec{x}}\frac{1}{|\vec{x}-\vec{y}|}dS_{\vec{x}}=0~,
  \label{eq:6}
\end{equation}
which can easily be proven with the divergence theorem and using $\Delta_{\vec x} \frac{1}{|\vec{x} - \vec{y}|}=0$. For taking the limit $\vec{y}\to \partial \Omega$ we will split the integration domain into two parts. One integral is for all surface except for the excluded region with a radius $\varepsilon$ centred at $\vec{y}$ which in the limit $\varepsilon \to 0$ gives us a principal value integral. The other one is around the half sphere whose value can be calculated analytically
\begin{equation}
  \int_{\partial \Omega_{\varepsilon}} \vec{n}_{\vec{x}} \bcdot \nabla_{\vec{x}} \frac{1}{|\vec{x} - \vec{y}|} dS_{\vec{x}} = \int_{\partial \Omega_{\varepsilon}} \frac{1}{|\vec{x} - \vec{y}|^2} dS_{\vec{x}} = 2 \pi \int_{\pi/2}^{\pi} \sin \theta d \theta = 2 \pi~.
\end{equation}
Taking the limit $\varepsilon \to 0$ we conclude that limiting value of the integral \eqref{eq:6} is
\begin{equation}
  \dashint_{\partial \Omega} \vec{n}_{\vec{x}} \bcdot \nabla_{\vec{x}} \frac{1}{|\vec{x} - \vec{y}|} dS_{\vec{x}} = \int_{\partial \Omega} - \int_{\partial \Omega_{\varepsilon}} = -2 \pi~,
  \label{eq:18}
\end{equation}
which we use for the singularity subtraction in \eqref{eq:Potential-BIE-regular}.

When $\vec{y} \ni \Omega$ we have another identity
\begin{equation}
  \int_{\partial \Omega} \vec{n}_{\vec{x}} \times \nabla_{\vec{x}} \frac{1}{|\vec{x} - \vec{y}|} dS_{\vec{x}} = \int_{ \Omega} \nabla_{\vec{x}} \times \nabla_{\vec{x}} \frac{1}{|\vec{x} - \vec{y}|} dV_{\vec{x}} = 0 
\end{equation}
since the curl of a potential field is $0$. For taking the limit $\vec{y} \to \vec{x}$ we again split the integration domain into two parts. Integral around the  half sphere is $0$ as the normal vector $\vec{n}_{\vec x}$ is collinear with $\nabla_{\vec{x}} \frac{1}{|\vec{x} - \vec{y}|}$ when $\vec y$ is at the center of half sphere. So we conclude that the principal value integral when $\vec y \in \partial \Omega$ is
\begin{equation}
  \dashint_{\partial \Omega} \vec{n}_{\vec{x}} \times \nabla_{\vec{x}} \frac{1}{|\vec{x} - \vec{y}|} dS_{\vec{x}} = 0.
  \label{eq:17}
\end{equation}

For deriving the formula used in \eqref{eq:Biot-Savart-integral-regular} we start with the identity
\begin{equation}
  \vec{A} \times (\vec{B} \times \vec{C}) = \vec{A} (\vec{B} \bcdot \vec{C}) - \vec{C} (\vec{A} \bcdot \vec{B}) + \vec{C} \times (\vec{B} \times \vec{A})~.
\end{equation}
Putting in $\vec A \to \nabla_{\vec{x}} \frac{1}{|\vec{x} - \vec{y}|}$, $\vec{B} \to \vec{n}_{\vec{x}}$, $\vec{C} \to \rm{P} \vec{H}_{\vec{x}}$ gives us the following integral identity
\begin{eqnarray}
  \dashint_{\partial \Omega} \nabla_{\vec{x}} \frac{1}{|\vec{x} - \vec{y}|} \times (\vec{n}_{\vec{x}} \times \vec{H}_{\vec{x}}) dS_{\vec{x}} = \dashint_{\partial \Omega} \rm{P} \vec{H}_{\vec{x}} \times (\vec{n}_{\vec{x}} \times \nabla_{\vec{x}} \frac{1}{|\vec{x} - \vec{y}|} ) dS_{\vec{x}} \nonumber \\
  -
  \dashint_{\partial \Omega} \rm{P} \vec{H}_{\vec{x}} \left(
  \vec{n}_{\vec{x}} \bcdot \nabla_{\vec{x}} \frac{1}{|\vec{x} - \vec{y}|}
  \right)
  dS_{\vec{x}}~.
\end{eqnarray}
Using the singularity subtraction identities \eqref{eq:17} and \eqref{eq:18} for both integrals on the right side we obtain
\begin{eqnarray}
  \dashint_{\partial \Omega} \nabla_{\vec{x}} \frac{1}{|\vec{x} - \vec{y}|} \times (\vec{n}_{\vec{x}} \times \vec{H}_{\vec{x}}) dS_{\vec{x}} = \dashint_{\partial \Omega} (\rm{P} \vec{H}_{\vec{x}} - \rm{P} \vec{H}_{\vec{y}}) \times (\vec{n}_{\vec{x}} \times \nabla_{\vec{x}} \frac{1}{|\vec{x} - \vec{y}|} ) dS_{\vec{x}} \nonumber \\
  -
  \dashint_{\partial \Omega} (\rm{P} \vec{H}_{\vec{x}} - \rm{P} \vec{H}_{\vec{y}}) \left(
  \vec{n}_{\vec{x}} \bcdot \nabla_{\vec{x}} \frac{1}{|\vec{x} - \vec{y}|}
  \right)
  dS_{\vec{x}} - 2 \pi \rm{P} \vec{H}_{\vec{y}}.
\end{eqnarray}
The right side is now weakly singular and thus can be calculated by means of trapezoidal quadrature for a nonsingular integral and using a transformation of variables for weakly singular elements.

\section{Identity\label{sec:identity}}

To transform the integral
\begin{equation}
\int \nabla_{\vec{x}}\bcdot\Bigl(\frac{\vec{n}_{\vec{x}}}{|\vec{y}-\vec{x}|}+\frac{(\vec{y}-\vec{x})(\vec{n}_{\vec{x}}\bcdot(\vec{y}-\vec{x}))}{|\vec{y}-\vec{x}|^{3}}\Bigr)dS_{\vec{x}}
\label{Eq:C1}
\end{equation}
we consider the two terms in (\ref{Eq:C1}) separately.
\begin{equation}
\vec{I}^{(1)}=\int \frac{\nabla_{\vec{x}}\bcdot\vec{n}_{\vec{x}}}{|\vec{y}-\vec{x}|}dS_{\vec{x}}~.
\label{Eq:C2}
\end{equation}
Multiplying (\ref{Eq:C2}) by an arbitrary constant vector $\vec{a}$ and using the relation
$$
\vec{a}\nabla\bcdot\vec{n}=\nabla\times[\vec{a}\times\vec{n}]+(\vec{a}\bcdot\nabla)\vec{n}
$$
we have
\begin{equation}
\vec{a}\bcdot\vec{I}^{(1)}=\int\frac{\vec{n}_{\vec{x}}\bcdot\nabla_{\vec{x}}\times[\vec{a}\times\vec{n}_{\vec{x}}]}{|\vec{y}-\vec{x}|}dS_{\vec{x}}+\int\frac{\vec{n}_{\vec{x}}\bcdot(\vec{a}\bcdot\nabla_{\vec{x}})\vec{n}_{\vec{x}}}{|\vec{y}-\vec{x}|}dS_{\vec{x}}~.
\label{Eq:C3}
\end{equation}
Last term on the right-hand side of (\ref{Eq:C3}) is zero since
$$
\vec{n}\bcdot(\vec{a}\bcdot\nabla)\vec{n}=\frac{1}{2}(\vec{a}\bcdot\nabla)\vec{n}^{2}=0~.
$$
The first term on the right-hand side is transformed as follows
\begin{equation}
\int\frac{\vec{n}_{\vec{x}}\bcdot\nabla_{\vec{x}}\times[\vec{a}\times\vec{n}_{\vec{x}}]}{|\vec{y}-\vec{x}|}dS_{\vec{x}}=
\int\vec{n}_{\vec{x}}\bcdot\nabla_{\vec{x}}\times\Bigl[\frac{\vec{a}\times\vec{n}_{\vec{x}}}{|\vec{y}-\vec{x}|}\Bigr]dS_{\vec{x}}-\int\vec{n}_{\vec{x}}\bcdot\Bigl[\nabla_{\vec{x}}\frac{1}{|\vec{y}-\vec{x}|}\times[\vec{a}\times\vec{n}_{\vec{x}}]\Bigr]dS_{\vec{x}}~.
\label{Eq:C4}
\end{equation}
The first term on the right-hand side of relation (\ref{Eq:C4}) is zero since the integration is carried out over a closed surface.
Since the vector $\vec{a}$ is arbitrary we obtain
\begin{equation}
\vec{I}^{(1)}=-\int\vec{n}_{\vec{x}}(\nabla_{\vec{x}}\frac{1}{|\vec{y}-\vec{x}|}\bcdot\vec{n}_{\vec{x}})dS_{\vec{x}}+\int\frac{(\vec{y}-\vec{x})}{|\vec{y}-\vec{x}|^{3}}dS_{\vec{x}}~.
\label{Eq:C5}
\end{equation}
The second term in (\ref{Eq:C1}) is transformed similarly
\begin{equation}
\vec{I}^{(2)}=\int\nabla_{\vec{x}}\bcdot\vec{n}_{\vec{x}}\frac{(\vec{y}-\vec{x})(\vec{n}_{\vec{x}}\bcdot(\vec{y}-\vec{x}))}{|\vec{y}-\vec{x}|^{3}}dS_{\vec{x}}~;
\label{Eq:C6}
\end{equation}
\begin{equation}
\vec{a}\bcdot\vec{I}^{(2)}=\int\nabla_{\vec{x}}\bcdot\vec{n}_{\vec{x}}\frac{\vec{a}\bcdot(\vec{y}-\vec{x})(\vec{n}_{\vec{x}}\bcdot(\vec{y}-\vec{x}))}{|\vec{y}-\vec{x}|^{3}}dS_{\vec{x}}~.
\label{Eq:C7}
\end{equation}
Using the relation
$$
\vec{n}_{\vec{x}}\bcdot(\vec{y}-\vec{x})\nabla_{\vec{x}}\cdot\vec{n}_{\vec{x}}=-2+\vec{n}_{\vec{x}}\bcdot\nabla_{\vec{x}}\times[(\vec{y}-\vec{x})\times\vec{n}_{\vec{x}}]
$$
we obtain
\begin{equation}
\vec{a}\bcdot\vec{I}^{(2)}=\int\frac{\vec{a}\bcdot(\vec{y}-\vec{x})}{|\vec{y}-\vec{x}|^{3}}\vec{n}_{\vec{x}}\bcdot\nabla_{\vec{x}}\times[(\vec{y}-\vec{x})\times\vec{n}_{\vec{x}}]]dS_{\vec{x}}-2\int\frac{\vec{a}\bcdot(\vec{y}-\vec{x})}{|\vec{y}-\vec{x}|^{3}}dS_{\vec{x}}~.
\label{Eq:C8}
\end{equation}
The first term on the right-hand side of (\ref{Eq:C8}) may be transformed as follows
\begin{equation}
\int\vec{n}_{\vec{x}}\bcdot\nabla_{\vec{x}}\times\Bigl[\frac{\vec{a}\bcdot(\vec{y}-\vec{x})}{|\vec{y}-\vec{x}|^{3}}(\vec{y}-\vec{x})\times\vec{n}_{\vec{x}}\Bigr]dS_{\vec{x}}-\int\vec{n}_{\vec{x}}\bcdot\nabla_{\vec{x}}\Bigl[\frac{\vec{a}\bcdot(\vec{y}-\vec{x})}{|\vec{y}-\vec{x}|^{3}}\times[(\vec{y}-\vec{x})\times\vec{n}_{\vec{x}}]\Bigr]dS_{\vec{x}}
\label{Eq:C9}
\end{equation}
The first term on the right-hand side of relation (\ref{Eq:C9}) is  zero since the integration is carried out over the closed surface. The second term on the right-hand side of relation (\ref{Eq:C9}) equals
\begin{equation}
\frac{\vec{n}_{\vec{x}}\bcdot(\vec{y}-\vec{x})\vec{a}\bcdot\vec{n}_{\vec{x}}}{|\vec{y}-\vec{x}|^{3}}+2\frac{\vec{a}\bcdot(\vec{y}-\vec{x})}{|\vec{y}-\vec{x}|^{3}}-3\frac{\vec{a}\bcdot(\vec{y}-\vec{x})(\vec{n}_{\vec{x}}\bcdot(\vec{y}-\vec{x}))^{2}}{|\vec{y}-\vec{x}|^{5}}~.
\label{Eq:C10}
\end{equation}
As a result for $\vec{I}$ we have
\begin{equation}
\vec{I}=\vec{I}^{(1)}+\vec{I}^{(2)}=\int\frac{\vec{y}-\vec{x}}{|\vec{y}-\vec{x}|^{3}}dS_{\vec{x}}-3\int\frac{(\vec{y}-\vec{x})\vec{n}_{\vec{x}}\bcdot(\vec{y}-\vec{x}))^{2}}{|\vec{y}-\vec{x}|^{5}}dS_{\vec{x}}~.
\label{Eq:C11}
\end{equation}
\begin{eqnarray}
\frac{\vec{n}_{\vec{y}}\bcdot(\vec{y}-\vec{x})}{|\vec{y}-\vec{x}|^{3}}-3\frac{((\vec{y}-\vec{x})\bcdot\vec{n}_{\vec{x}})^{2}\vec{n}_{\vec{y}}\bcdot(\vec{y}-\vec{x})}{|\vec{y}-\vec{x}|^{5}}-\\ \nonumber
\frac{(\vec{y}-\vec{x})(\vec{n}_{\vec{x}}+\vec{n}_{\vec{y}})}{\vec{y}-\vec{x}|^{3}}\Bigl(1-3\frac{\vec{y}-\vec{x})\bcdot\vec{n}_{\vec{x}}(\vec{y}-\vec{x})\bcdot\vec{n}_{\vec{y}}}{|\vec{y}-\vec{x}|^{2}}\Bigr)=\\ \nonumber
-\frac{(\vec{y}-\vec{x})\cdot\vec{n}_{\vec{x}}}{|\vec{y}-\vec{x}|^{3}}+3\frac{(\vec{y}-\vec{x})\bcdot\vec{n}_{\vec{x}}(\vec{n}_{\vec{y}}\bcdot(\vec{y}-\vec{x})^{2}}{|\vec{y}-\vec{x}|^{5}}
\end{eqnarray}
Since the integral of right-hand side equals zero we arrive at the identity used in the text
\begin{eqnarray}
\vec{n}_{\vec{y}}\int\Bigl(\frac{1}{R_{1}}+\frac{1}{R_{2}}\Bigr)\Bigl(\frac{\vec{n}_{\vec{x}}}{|\vec{y}-\vec{x}|}+\frac{(\vec{y}-\vec{x})(\vec{y}-\vec{x})\bcdot\vec{n}_{\vec{x}}}{|\vec{y}-\vec{x}|^{3}}\Bigr)dS_{\vec{x}}=\\ \nonumber
\int\frac{(\vec{y}-\vec{x})(\vec{n}_{\vec{x}}+\vec{n}_{\vec{y}})}{|\vec{y}-\vec{x}|^{3}}\Bigl(1-3\frac{(\vec{y}-\vec{x})\bcdot\vec{n}_{\vec{x}}(\vec{y}-\vec{x})\bcdot\vec{n}_{\vec{y}}}{|\vec{y}-\vec{x}|^{2}}\Bigr)
\end{eqnarray}

\bibliographystyle{jfm}

\bibliography{bib2}

\end{document}